\documentclass[a4paper]{IEEEtran}
\usepackage[T1]{fontenc}
\usepackage[utf8]{inputenc}
\usepackage{subfig}
\usepackage{pgfplots}
\usepackage{tikz}
\usepackage{paralist}
\usepackage{amsmath}
\usepackage{amssymb}
\usepackage{algorithm}
\usepackage[noend]{algorithmic}
\usepackage{mathtools}

\usetikzlibrary{patterns}
\usetikzlibrary{positioning}
\usetikzlibrary{topaths}

\pgfplotsset{compat = 1.3}

\DeclarePairedDelimiter{\ceil}{\lceil}{\rceil}

\pgfplotsset{
  mybars/.style = {
    error bars/error mark = {none},
    error bars/error bar style = {solid, #1, double distance = -0.4 pt},
    error bars/y dir = both,
    error bars/y explicit
  }
}

\tikzstyle{node} = [circle, draw, minimum size = 0.6 cm]

\DeclareMathOperator{\AU}{AU}
\DeclareMathOperator{\CU}{CU}
\DeclareMathOperator{\cost}{cost}
\DeclareMathOperator{\decide}{decide}
\DeclareMathOperator{\edge}{edge}
\DeclareMathOperator{\myrelax}{relax}
\DeclareMathOperator{\target}{target}
\DeclareMathOperator{\trace}{trace}
\DeclareMathOperator{\Poisson}{Poisson}

\newcommand{\twodots}{\mathinner {\ldotp \ldotp}}
\newcommand{\CUO}[2]{[#1\twodots#2]}
\newcommand{\ALL}{\Omega}
\newcommand{\eattrs}[2]{(#1, #2)}
\newcommand{\lbl}[3]{(#1, #2, #3)}
\newcommand{\nulledge}{e_\emptyset}

\title{Generic Dijkstra for Optical Networks}

\author{Ireneusz Szcześniak, Andrzej Jajszczyk, Bożena Woźna-Szcześniak

  \thanks{This work was supported in part by the Regional Initiative
    of Excellence project number 020/RID/2018/19 of the Polish
    Ministry of Science and Higher Education, and in part by the
    postdoctoral fellowship number DEC-2013/08/S/ST7/00576 of the
    Polish National Science Centre.  The simulation results were
    obtained using PL-Grid, the Polish supercomputing infrastructure.
    This is the extended version of a conference paper
    \cite{10.1109/ONDM.2016.7494087}.}%

  \thanks{I.~Szcześniak is with the Institute of Computer and
    Information Sciences of the Częstochowa University of Technology,
    Częstochowa, Poland.}%
  \thanks{A.~Jajszczyk is with the Department of Telecommunications of
    the AGH University of Science and Technology, Kraków, Poland.}
  \thanks{B.~Woźna-Szcześniak is with the Institute of Mathematics and
    Computer Science of the Jan Długosz University, Częstochowa,
    Poland.}}

\begin{document}

\maketitle

\begin{abstract}
  We present the \emph{generic} Dijkstra shortest path algorithm: an
  efficient algorithm for finding a shortest path in an optical
  network, both in a wavelength-division multiplexed network, and an
  elastic optical network (EON).  The proposed algorithm is an enabler
  of real-time softwarized control of large-scale networks, and is not
  limited to optical networks.  The Dijkstra algorithm is a
  generalization of the breadth-first search, and we generalize the
  Dijkstra algorithm further to resolve the continuity and contiguity
  constraints of the frequency slot units required in EONs.
  Specifically, we generalize the notion of a label, change what we
  iterate with, and reformulate the edge relaxation so that vertices
  are revisited, loops avoided, and worse labels discarded.  We also
  used the typical constriction during edge relaxation to take care of
  the signal modulation constraints.  The algorithm can be used with
  various spectrum allocation policies.  We motivate and discuss the
  algorithm design, and provide our free, reliable, and generic
  implementation using the Boost Graph Library.  We carried out 85000
  simulation runs for realistic and random networks (Gabriel graphs)
  of 75 vertices with about a billion shortest-path searches, and
  found that the proposed algorithm outperforms considerably three
  other competing optimal algorithms that are frequently used.
\end{abstract}

\begin{IEEEkeywords}
  Dijkstra algorithm, shortest path routing, elastic optical network,
  wavelength-division multiplexing
\end{IEEEkeywords}

%%%%%%%%%%%%%%%%%%%%%%%%%%%%%%%%%%%%%%%%%%%%%%%%%%%%%%%%%%%%%%%%%%%%%%%%%%%

\section{Introduction}
\label{introduction}

% We start with routing, because that's what the paper is about.

\emph{Routing of a single connection in an optical network} is one of
the most important tasks of operating an optical network, and one of
many research problems of the optical network design, planning, and
operation.  In the wavelength-division multiplexed (WDM) network, the
problem is called the routing and wavelength assignment (RWA) problem,
and in the elastic optical network it is called the routing and
spectrum assignment (RSA) problem, or the routing, modulation, and
spectrum assignment (RMSA), if we take into account the constraints of
the signal modulation.  Related problems exist in other optical
networks: in the optical transport network (OTN) for the virtual and
contiguous concatenation, or in the space division multiplexing (SDM)
network for the fiber-core assignment.

% On EONs, and why slots are better than wavelengths.

In EONs, the optical spectrum (the erbium window) is divided into fine
\emph{frequency slot units} (of, e.g., 6.25 GHz width), or just
\emph{units}, as opposed to coarse fixed-grid channels (of, e.g., 25
GHz width) of WDM networks.  In EONs, \emph{contiguous units} are
concatenated to form a \emph{slot}.  Slots are tailored for a specific
demand, unlike WDM channels, thus making EONs more spectrum efficient
than WDM networks.

% Find an optimal path, and do it FAST.

Future optical networks should deal with dynamic traffic, where
connections are frequently established, and are of short duration as
opposed to the quasi-static WDM connections that are characteristic of
traditional networks.  Furthermore, given the increasing deployment of
optical networks, the network densification, the softwarization of the
network control, the ever-increasing need for bandwidth and agility,
further increased by the content-oriented services, network and
service orchestration, and the next generation wireless network
requirements, a shortest optical path should be found \emph{fast}.
The proposed algorithm enables real-time control of future optical
networks.

% Dynamic RWA, RSA and RMSA are not NP-hard.

The RWA, RSA and RMSA problems come in many versions, most notably
static (a.k.a.~offline) and dynamic (a.k.a.~online).  The objective of
the static version is to route a number of demands along shortest
paths in an unloaded network using the least spectrum.  The objective
of the dynamic version is to route a single demand along a shortest
path in a loaded network using the available spectrum.  The static
version is nondeterministic polynomial time complete (NP-complete),
but the dynamic version is not, because it can be solved tractably
(though inefficiently) by finding a shortest path in a number of
filtered graphs.

% Contribution in a nutshell: algorithm, simulations, implementation.

Our novel contribution is the algorithm which efficiently solves the
dynamic RWA, RSA and RMSA problems.  The algorithm is the
generalization of the Dijkstra shortest-path algorithm.  With
simulations, we demonstrate its efficiency in comparison to three
other optimal algorithms frequently used in research.  The
implementation of the algorithm using the Boost Graph Library is
available at \cite{gdwebsite}.  We published the first implementation
of the algorithm in 2013 \cite{sdiwebsite}.

% About Dijkstra

The shortest path Dijkstra algorithm is a premier graph algorithm,
amenable to various adaptations due to its simple and clever design.
Dijkstra is optimal (i.e., it finds a shortest path) and efficient (no
better algorithm has been proposed yet after half a century since it
was proposed), and follows the \emph{label-setting} paradigm, as
opposed to the label-correcting paradigm \cite{networkflows}.  At
first look, our generalization seems to discard the label-setting
paradigm in favor of the label-correcting paradigm, because we allow
for revisiting vertices, which Dijkstra does not do, and which is a
hallmark of the label-correcting algorithms.  But this is not so; the
proposed algorithm is still a label-setting algorithm, only with a
generalized notion of a label and reformulated edge relaxation.

% Paper organization.

The paper is organized as follows.  In Section \ref{related} we review
related works, in Section \ref{statement} we define the research
problem, in Section \ref{algorithm} we describe the proposed
algorithm, and in Section \ref{simulations} we report on the
simulation results.  Finally, Section \ref{conclusion} concludes the
paper.

%%%%%%%%%%%%%%%%%%%%%%%%%%%%%%%%%%%%%%%%%%%%%%%%%%%%%%%%%%%%%%%%%%%%%%%%%%%

\section{Related works}
\label{related}

% The plan for this section is:
% - this article is an extension,
% - dynamic RSA was a problem,
% - introduce the three other routing algorithms:
%   * the filtered-graph algorithm,
%   * the brute-force algorithm,
%   * the unlimited Yen KSP,
% - mention some other algorithms,
% - conclude with why our algorithm doesn't fail, while some other do.

% The main differences with the conference paper.

We extended our conference paper \cite{10.1109/ONDM.2016.7494087} in a
number of significant ways.  First, we refined the algorithm
description by defining better the relation between labels.  Second,
we took into account the signal modulation constraints.  Third, since
the proposed algorithm is not heuristic, we evaluated its performance
in comparison with three other optimal algorithms, and not with
heuristic algorithms as before.

% Generic is the name.

The proposed algorithm was inspired by the \emph{generic programming}
paradigm which is based on mathematical abstraction: generic data
structures and algorithms can operate on any algebraic structure,
provided it has the required properties, such as operations or
ordering relations \cite{generic}.  We call the proposed algorithm the
\emph{generic Dijkstra} algorithm as a tribute to generic programming.

% The dynamic problem definition.

The dynamic RWA, RSA, and RMSA problems are defined in the literature
in two different ways.  First, they can be defined to optimize the
overall network performance expressed in, e.g., the bandwidth blocking
probability for a set of demands, bearing similarity to the static
version, thus this problem is considered hard \cite{Zang00areview}.
Second, these problems can be defined to optimize a single connection
(i.e., to find a shortest path or a path of minimal cost), since in
an operational network requests usually arrive sequentially
\cite{10.1016/j.osn.2014.02.003}.  We concentrate on the second
definition of the dynamic RWA, RSA, and RMSA problems, i.e., we
optimize a single connection.

% The proofs.

The static RWA was proven to be NP-complete nearly three decades ago,
and then the static RSA was proven NP-complete too
\cite{10.1109/INFCOM.2011.5934939}.  The status of the
single-connection dynamic RWA, RSA, and RMSA problems was unclear: no
proof of NP-completeness was proposed, but heuristic algorithms and
linear-programming formulations were proposed and reviewed
\cite{10.1109/COMST.2015.2431731, 10.1016/j.osn.2016.08.003,
10.1007/s11107-017-0700-5}.

% The filtered-graphs algorithm, and the complexity upper bound.

The dynamic RWA, RSA, and RMSA problems can be solved inefficiently by
finding shortest paths in filtered graphs.  A filtered graph retains
only those edges which can support a given slot.  For a given demand
and available modulations, we compute set $S$ of slots for which we
filter the input graph, and search for a shortest path.  From among
the shortest paths found, the shortest one is selected.  This
algorithm, termed \emph{filtered-graphs algorithm}, is of
$O(|S| \times |V| \log |V|)$ complexity, where $V$ is the set of
vertices of a (sparse, we assume) graph.  This computational
complexity is the upper bound, and a proof that these problems are
tractable.

% We do better.

We argue that the filtered-graphs algorithm is inefficient and we show
through simulations on realistic networks that the proposed algorithm
is considerably faster than the filtered-graphs algorithm.  To the
best of our knowledge, we are the first to propose an optimal and
efficient algorithm for the dynamic routing problem, as simulated in
realistic optical networks.

% The Velasco claim.

In \cite{10.1007/s11107-012-0378-7} the authors report that by
computing in advance the set of slots that can be assigned to a
demand, the complexity added by the contiguity constraint is removed.
We, in contrast, show that the time performance of the filtered-graphs
algorithm is worse than the time performance of the proposed
algorithm, because we process not a slot, but contiguous units, which
can include many slots.

% The brute-force algorithm.

In \cite{10.1364/JOCN.6.001115} the authors proposed an algorithm for
solving the dynamic RSA problem with the brute-force search strategy
of enumerating the paths capable of supporting a demand.  The
algorithm does not use the edge relaxation to limit the search space.
The algorithm stores the complete paths in a priority queue, and does
not use the dynamic-programming principle of reusing intermediate
results, like Dijkstra does with node labels.  For these reasons, we
refer to this algorithm as the \emph{brute-force algorithm}.

% The unlimited and limited Yen KSP.

In \cite{10.1364/JOCN.4.000603} the authors proposed an algorithm,
which checks whether the consecutive paths provided by the Yen
$K$-shortest path (KSP) algorithm can support a demand.  When $K$ is
limited to some value (e.g., $K = 10$), we call this algorithm the
\emph{limited Yen KSP}, and if there is no limit on $K$, we call the
algorithm the \emph{unlimited Yen KSP}.  The limited Yen KSP algorithm
is heuristic, while the unlimited Yen KSP is optimal.  If there is a
path capable of supporting the demand, the unlimited Yen KSP algorithm
will eventually find this path, but its $K$ could be very large,
possibly a million, even for small graphs.

% The other algorithm, and its shortcommings:
%
% - storing the complete paths,
%
% - the domination relation,
%
% - a path cost vector stores the info for all wavelengths,
%
% - non-dominant paths can have overlapping wavelengths, and that
%   leads to the explosion of the search space,
%
% - the problematic two phases.

% The other algorithms, and the relations.

In \cite{10.1109/TNET.2011.2138717}, the authors proposed a dynamic
RWA algorithm, which is based on the algorithm in
\cite{10.1016/j.comnet.2008.06.016}, and in turn based on the
Dijkstra-based algorithm proposed in \cite{gutierrez}.  In
\cite{gutierrez}, the authors introduced the transitive path
domination relation to solve the max-min problem, where the network
resources are continuous.  This relation was adapted in
\cite{10.1016/j.comnet.2008.06.016} for discrete network resources and
then used \cite{10.1109/TNET.2011.2138717} to model the availability
of wavelengths.  That adaptation, however, leads to a less efficient
search, since it was defined with the $\ge$ relation for the
wavelength availability vector, which models the $\supseteq$ set
relation.  We, in contrast, define the relation of the incomparability
of solution labels using the $\supsetneq$ relation for a set of
contiguous units, and show that this relation is intransitive.

% The path cost vector problem.

The authors in \cite{10.1109/TNET.2011.2138717}, and
\cite{10.1016/j.comnet.2008.06.016} report on the complexity of their
algorithms, which may be exponential.  This complexity is the result
of the way the path cost vectors are defined, compared, and produced
when an edge is traversed.  In \cite{10.1109/TNET.2011.2138717} the
path cost vector stores the availability information of all
wavelengths of a path.  The edge traversal produces paths which are
non-dominant, but which can have overlapping wavelengths, thus
possibly leading to the explosion of the search space (the set of
non-dominant paths).  We, in contrast, store in a solution label a
single sequence of contiguous units (represented by two integer
numbers), which is faster to process.  During the edge relaxation, we
produce a set of incomparable labels (without overlapping units),
which limits the search space (the set of incomparable labels).

% We have one phase only.

The algorithm in \cite{10.1109/TNET.2011.2138717} finds an optimal
solution in two phases: in the first phase, the complete set of
non-dominated solutions is obtained, and in the second phase an
optimal solution is selected.  In contrast, our algorithm can make
optimization decisions for intermediate solutions by sorting the
priority queue elements, taking into account, e.g., the spectrum
allocation policy.  Thus, our algorithm has a single phase, and
produces a single result, which is more efficient.

% Modified Dijkstra.

In \cite{10.1364/JOCN.4.000603}, the authors also proposed a heuristic
algorithm, termed a modified Dijkstra algorithm, which is the Dijkstra
algorithm with the constrained edge relaxation, where a candidate path
is rejected if it cannot support a demand.

% One-stage constrained Yen.

In \cite{10.1109/JLT.2014.2315041}, the authors proposed a heuristic
algorithm, which is a constrained Yen $K$-shortest path algorithm that
drops the path deviations incapable of supporting a demand.  The Yen
algorithm delegates the shortest path search to the Dijkstra
algorithm.

% Conclusion: our algorithm is awesome.

The limited Yen, the modified Dijkstra, and the constrained Yen
algorithms, which find a candidate path with the Dijkstra algorithm,
fail to find a shortest path meeting the spectrum and modulation
constraints, when there is a shorter path unable to meet the spectrum
and modulation constraints, because that shorter path diverts the
algorithms into a dead end.

%%%%%%%%%%%%%%%%%%%%%%%%%%%%%%%%%%%%%%%%%%%%%%%%%%%%%%%%%%%%%%%%%%%%%%%%%%%

\section{Problem statement}
\label{statement}

% What we are given.

\noindent{}Given:

\begin{itemize}

\item directed multigraph $G = (V, E)$, where $V = \{v_i\}$ is a set
  of vertices, and $E = \{e_i\}$ is a set of edges,

\item cost function $\cost(e_i)$, which gives non-negative cost
  (length) of edge $e_i$,

\item available units function $\AU(e_i)$, which gives the set of
  available units of edge $e_i$, which do not have to be contiguous,

\item $s$ and $t$ are the source and target vertices of the demand,

\item a \emph{decision function of monotonically increasing
  requirements}, which returns true if a candidate solution (the given
  contiguous units at the given cost) can support the demand,
  otherwise false,

\item the set of all units $\Omega$ on every edge.

\end{itemize}

% What we are looking for.

\noindent{}Find:

\begin{itemize}

\item a shortest path (a sequence of edges),

\item continuous and contiguous units.

\end{itemize}

% What a CU is.

We refer to a set of contiguous units as a CU.  We denote a CU with
the units starting at $a$ and ending at $b$ inclusive as $\CUO{a}{b}$.
For instance, $\CUO{0}{2}$ denotes units 0, 1 and 2.  A set of units
can be treated as a set of CUs.  For instance, $\{0, 1, 3, 4, 5\}$ and
$\{\CUO{0}{1}, \CUO{3}{5}\}$ are the same.

% Incomparability of CUs.

Two CUs are \emph{incomparable}, when one is not included in the
other.  For instance, $\CUO{0}{2}$ is incomparable with $\CUO{2}{3}$.
We denote the incomparability of CUs with the $\parallel$ relation.
For instance, $\CUO{0}{2} \parallel \CUO{2}{3}$.

% The decision function, and why it's defined that way.

We intentionally stated the problem generically by introducing the
decision function to consider the RWA, RSA, and RMSA problems at once.
The decision function is responsible for accepting or rejecting a
candidate solution, and gives a user leeway to define what an
acceptable candidate solution is.  For RWA, the function should check
whether a CU has at least one unit (wavelength), for RSA, whether a CU
has at least the number of contiguous units required by the demand,
and for RMSA, whether a CU has at least the number of contiguous units
required by the demand at a given cost (distance).  The decision
function could also check other parameters such as the signal quality.

% The monotonically increasing requirements of the function.

The requirements of the decision function should be monotonically
increasing as the cost increases, i.e., a candidate solution rejected
at a given cost could not be accepted at a higher cost.  It is a valid
assumption, since as the distance grows, the number of required
contiguous units can only grow.  This assumption allows us to reject a
candidate solution due to its insufficient number of the available
contiguous units, because as distance grows, that rejected solution
would fail to provide the same or greater number of required
contiguous units anyway.

% Why is the bitrate not mentioned.

The demand bitrate is not a given of the stated problem, because that
would narrow the problem statement.  If need be, the demand bitrate
should be considered by the decision function.  In Section
\ref{simulations} we use the proposed algorithm to solve the RMSA
problem, and there, in Algorithm \ref{a:decide}, we define a decision
function which takes into account the demand bitrate.

%%%%%%%%%%%%%%%%%%%%%%%%%%%%%%%%%%%%%%%%%%%%%%%%%%%%%%%%%%%%%%%%%%%%%%%%%%%

% Presentation plan:
%
% * make the three observations,
%
% * present the three changes,
%
% * mention the constriction, almost in passing,
%
% * discuss the algorithm.
%
% At each of these stages give away the details in doses, don't give
% them all at once, so that the explanation is clear and
% comprehensive.

\section{Proposed algorithm}
\label{algorithm}

% Generalization and constriction.

We generalize and constrain the shortest path Dijkstra algorithm to
find a shortest path in EONs for a given demand.  The generalization
is novel, and the constriction is trivial.  The generalization
resolves the unit continuity and contiguity constraints, while the
constriction takes into account the signal modulation constraints.

% Tutorial: the label in the label-setting algorithms.

In label-setting algorithms, a label is associated with a vertex, and
gives information on what cost and how to reach that vertex from the
source.  In Dijkstra, the label is defined as a pair of a cost and a
preceding vertex.  Each vertex has at most one label, which we call
the \emph{vertex label}.  For a given vertex, the vertex label is
initially \emph{tentative}, because it can be updated by the edge
relaxation, and then becomes \emph{permanent} when the vertex is
visited.

% Tutorial: Dijkstra is label-setting.

The Dijkstra algorithm is label setting in that once a vertex is
visited, its label is set for good (the status of the vertex label
changes from \emph{tentative} to \emph{permanent}), but before that
happens, the vertex label converges to its optimum by \emph{edge
relaxation}.

% Tutorial: the edge relaxation.

In Dijkstra, when relaxing edge $e$, a tentative vertex label is
updated with a better candidate label.  The tentative vertex label, if
it exists (i.e., it has been found by some earlier relaxation), is the
tentative vertex label of the target vertex of edge $e$.  The
candidate label is the label produced for edge $e$, and tells the cost
of reaching the target vertex of edge $e$.  The candidate label is
better than the tentative vertex label if it has a lower cost.

% This figure has to be here, so that it's placed where I want.

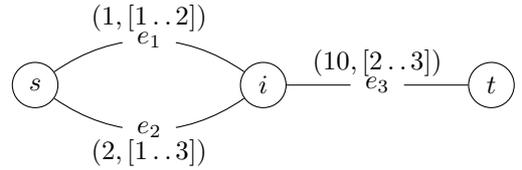
\begin{figure}
  \centering
  \begin{tikzpicture}[node distance = 3 cm]
    \node [node] (s) {$s$};
    \node [node, right of = s] (i) {$i$};
    \node [node, right of = i] (t) {$t$};

    \path (s) edge [bend left = 35]
    node [circle, fill = white] {$e_1$}
    node [above] {$\eattrs{1}{\CUO{1}{2}}$} (i);

    \path (s) edge [bend right = 35]
    node [circle, fill = white] {$e_2$}
    node [below] {$\eattrs{2}{\CUO{1}{3}}$} (i);

    \path (i) edge
    node [circle, fill = white] {$e_3$}
    node [above] {$\eattrs{10}{\CUO{2}{3}}$} (t);

  \end{tikzpicture}
  \caption{Example for vertex revisiting and looping.}
  \label{f:example1}
\end{figure}

\subsection{Observations on the stated problem}
\label{observations}

The following three observations shaped the generalization.

\subsubsection{Revisit vertices}

% In Dijkstra: one visit.  Not us.  No mention of CUs yet.

In Dijkstra, a vertex is visited only once for a single label.  We,
however, want to \emph{revisit} a vertex even for a label of a higher
cost than the vertex label, because it may eventually yield a shortest
path capable of supporting a given demand.

% Revisiting example - intro.

To demonstrate vertex revisiting, we show an example in
Fig.~\ref{f:example1}, where an edge is annotated with the length and
the available units, e.g., $\eattrs{1}{\CUO{1}{2}}$ says the edge is
of length 1 with units 1 and 2 available.  We are searching for a
shortest path with two units from vertex $s$ to vertex $t$.

% Revisiting example - description.

In the first iteration of Dijkstra, vertex $s$ is visited, vertex $i$
is discovered along edge $e_1$, and the discovery along edge $e_2$ is
discarded because of a higher cost.  In the second iteration, vertex
$i$ is visited, and now we know that we can get to vertex $i$ along
edge $e_1$ at cost 1, and with $\CUO{1}{2}$.  The problem is that
vertex $t$ cannot be discovered, because the spectrum continuity
constraint would be violated: the demand requires two units, vertex
$i$ is visited with $\CUO{1}{2}$, but $\AU(e_3) = \CUO{2}{3}$.  We
reach a dead end; the search stops with no solution.

% Revisiting example - conclusion.

Continuing with the example, and allowing for vertex revisiting, now
vertex $i$ is discovered along both parallel edges $e_1$ and $e_2$,
and none of the discoveries is discarded.  Now there are two tentative
labels for vertex $i$.  Then vertex $i$ is visited along edge $e_1$ at
cost 1 with $\CUO{1}{2}$, and then revisited along edge $e_2$ at cost
2 with $\CUO{1}{3}$, thus allowing vertex $t$ to be discovered, end
eventually visited at cost 12 with $\CUO{2}{3}$.

% This figure has to be here, so that it's placed where I want.

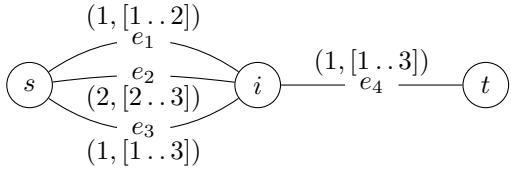
\begin{figure}
  \centering
  \begin{tikzpicture}[node distance = 3 cm]
    \node [node] (s) {$s$};
    \node [node, right of = s] (i) {$i$};
    \node [node, right of = i] (t) {$t$};

    \path (s) edge [bend left = 35]
    node [circle, fill = white] {$e_1$}
    node [above] {$\eattrs{1}{\CUO{1}{2}}$} (i);

    \path (s) edge [bend right = 35]
    node [circle, fill = white] {$e_3$}
    node [below] {$\eattrs{1}{\CUO{1}{3}}$} (i);

    \path (s) edge [bend left = 7.5]
    node [bend left = 20, circle, fill = white] {$e_2$}
    node [below] {$\eattrs{2}{\CUO{2}{3}}$} (i);

    \path (i) edge
    node [circle, fill = white] {$e_4$}
    node [above] {$\eattrs{1}{\CUO{1}{3}}$} (t);

  \end{tikzpicture}
  \caption{Example for discarding worse labels.}
  \label{f:example2}
\end{figure}

\subsubsection{Avoid loops}

% How loops are avoided in Dijkstra.

In Dijkstra, loops are avoided, because an edge is relaxed only if it
yields a candidate label of a lower cost.  Since edge costs are
non-negative, loops cannot decrease cost, and so they will not be
allowed by edge relaxation.

% We get loops because of revisiting.

The problem is we will find loops if we revisit vertices at higher
costs.  For instance, considering the same example in
Fig.~\ref{f:example1}: when we visit vertex $i$, we rediscover vertex
$s$ and later revisit it, thus finding the loop with edges $e_1$ and
$e_2$.

% Revisiting refined: take the CUs into account.

To avoid loops, we refine when we can revisit a vertex. We still allow
a revisit at a higher cost, \emph{but only for a CU not included in
the CUs of previous visits.}  Therefore, a vertex is visited and
possibly revisited always at the lowest cost for a CU not included in
the CUs of previous visits.

% Example: loop avoided.

For example, in Fig.~\ref{f:example1}, we start the search by visiting
vertex $s$ at cost 0, and with the CU of $\ALL$.  While visiting
vertex $i$ at cost 1, and with $\CUO{1}{2}$, we rediscover vertex $s$
along edge $e_2$, but the edge will not be relaxed, because vertex $s$
was already visited with $\ALL$, which includes $\CUO{1}{2}$.  This
example would hold even if the loop did not increase the cost.

% In Dijkstra: discard labels.

\subsubsection{Discard labels}

In Dijkstra, when a tentative vertex label is updated by the edge
relaxation, the previous value of the tentative label is discarded.
In generic Dijkstra, when we relax an edge, we can discard a number of
tentative labels.

% Example: discarding.

For instance, in Fig.~\ref{f:example2}, while visiting vertex $s$,
three edges are relaxed: first edge $e_1$ is relaxed at cost 1, and
with $\CUO{1}{2}$, next edge $e_2$ is relaxed at cost 2, and with
$\CUO{2}{3}$, and, finally, edge $e_3$ is relaxed at cost 1, and with
$\CUO{1}{3}$.  The results for edges $e_1$, and $e_2$ are discarded,
because the result for edge $e_3$ is better: at cost 1 it offers
$\CUO{1}{3}$, which includes both $\CUO{1}{2}$, and $\CUO{2}{3}$.
Thus, vertex $i$ is visited at cost 1, and with $\CUO{1}{3}$.

\subsection{Changes to the Dijkstra algorithm}
\label{changes}

Based on the observations, we motivate the changes to the Dijkstra
algorithm that make it generic, and applicable to the stated problem.

\subsubsection{Labels}

\begin{table*}
  \caption{Relations between labels $l_i$ and $l_j$ depending on their
    costs and CUs.}
  \label{t:relations}
  \centering
  \begin{tabular}{c|c|c|c|c|}
    \cline{2-5}
    & $\CU(l_i) \supsetneq \CU(l_j)$
    & $\CU(l_i) = \CU(l_j)$
    & $\CU(l_i) \subsetneq \CU(l_j)$
    & $\CU(l_i) \parallel \CU(l_j)$\\
    \hline
    \multicolumn{1}{|c|}{$\cost(l_i) < \cost(l_j)$}
    & $l_i < l_j$
    & $l_i < l_j$
    & $l_i \parallel l_j$
    & $l_i \parallel l_j$\\
    \hline
    \multicolumn{1}{|c|}{$\cost(l_i) = \cost(l_j)$}
    & $l_i < l_j$
    & $l_i = l_j$
    & $l_i > l_j$
    & $l_i \parallel l_j$\\
    \hline
    \multicolumn{1}{|c|}{$\cost(l_i) > \cost(l_j)$}
    & $l_i \parallel l_j$
    & $l_i > l_j$
    & $l_i > l_j$
    & $l_i \parallel l_j$\\
    \hline
  \end{tabular}
\end{table*}

% Our label.

We define a label as \emph{a tuple of cost, a CU, and a preceding
edge}, to keep track of the CU used.  For instance, label
$\lbl{1}{\CUO{1}{2}}{e_1}$ says that a vertex is reached at cost 1 and
with the CU of $\CUO{1}{2}$ along edge $e_1$.  To allow for
multigraphs, we keep a preceding edge, not a preceding vertex, in the
label.  The cost of label $l_i$ we denote as $\cost(l_i)$, and the
contiguous units as $\CU(l_i)$.  This label can be a candidate label,
a tentative vertex label, or a permanent vertex label.

% How we compare labels.

Label $l_i$ is better than label $l_j$ (or label $l_j$ is worse than
label $l_i$), denoted by $l_i < l_j$, if either:

\begin{enumerate}

  \item label $l_i$ offers a CU which includes the CU of label $l_j$
    at a lower cost than the cost of label $l_j$, i.e., $\cost(l_i) <
    \cost(l_j)$ and $\CU(l_i) \supset \CU(l_j)$, or

  \item label $l_i$ offers a CU which properly includes the CU of
    label $l_j$ at a cost that is lower than or equal to the cost of
    label $l_j$, i.e., $\cost(l_i) \le \cost(l_j)$ and $\CU(l_i)
    \supsetneq \CU(l_j)$.

\end{enumerate}

% Strict partial order.

Our $<$ label relation is a \emph{strict partial order}, since it is
irreflexive and transitive \cite{elements}.  Furthermore, in a strict
partial order some pairs can be incomparable.  We say that labels
$l_i$ and $l_j$ are incomparable, denoted by $l_i \parallel l_j$, when
neither $l_i < l_j$ nor $l_j < l_i$ holds.  Indeed, our labels can be
incomparable.

% Not strict weak order.

However, our $<$ order is not a \emph{strict weak order}, because the
incomparability of labels is not transitive.  For example, while
$\eattrs{0}{\CUO{1}{1}} \parallel \eattrs{2}{\CUO{1}{2}}$ and
$\eattrs{2}{\CUO{1}{2}} \parallel \eattrs{1}{\CUO{1}{1}}$ hold,
$\eattrs{0}{\CUO{1}{1}} \parallel \eattrs{1}{\CUO{1}{1}}$ does not,
because $\eattrs{0}{\CUO{1}{1}} < \eattrs{1}{\CUO{1}{1}}$ holds.

% The table of relations.

Table \ref{t:relations} shows the label relations depending on their
costs and CUs, where relation $l_i > l_j$ means $l_j < l_i$.

% Introduce the set of incomparable labels.

When we rediscover or revisit a vertex, we grow \emph{a set of
incomparable labels}, i.e., for any labels $l_i$ and $l_j$ that are
different, $l_i \parallel l_j$ is true, or equivalently $l_i < l_j$ is
false.  The incomparability of labels insures that in the set we do
not store a label that is worse than some other label.

\subsubsection{Iteration}

% In an interation, process a label, not a vertex.

In an iteration, Dijkstra processes a tentative vertex $v$ (i.e., a
vertex with a tentative label) of the lowest cost, while generic
Dijkstra processes a tentative label of the lowest cost, where the
edge of the tentative label has the target vertex $v$.  The labels we
iterate over are provided by the edge relaxation.

% Insert the label into the set of permanent incomparable labels.

An iteration corresponds to visiting (or revisiting) vertex $v$.  In
Dijkstra, only the status of the label of vertex $v$ changes from
tentative to permanent.  In generic Dijkstra, we insert the tentative
label into the set of permanent (optimal) incomparable labels of
vertex $v$.

\subsubsection{Relaxation}

% Reformulation of edge relaxation.

We reformulate the edge relaxation.  In Dijkstra, an edge is relaxed
when a candidate label is \emph{better} than the vertex label.  In
generic Dijkstra, we relax an edge when there is \emph{no better}
vertex label than the candidate label.  A small tweak.

% When the twist matters.

This tweak makes no difference when the relation between labels is a
strict total order, as in Dijkstra.  However, for our label order, the
tweak entails we relax an edge not only for better labels, but also
for incomparable labels.  A big difference.

% For a vertex we have sets of permanent and tentative labels.

In Dijkstra, a vertex has a single label, which is either tentative or
permanent.  In generic Dijkstra, a vertex has a set of tentative
labels, and a set of permanent labels.  Labels in these sets are
incomparable: no label in the two sets is better than some other
label, i.e., for any two different labels (tentative or permanent)
$l_i$ and $l_j$ of the given vertex, $l_i < l_j$ is false.  Our edge
relaxation maintains the labels in the two sets incomparable.

% Discarding labels.

As part of the relaxation we discard those tentative labels (of vertex
$v'$), which are worse than the candidate label.  The permanent labels
are left alone, because they are optimal.

\subsection{Constriction}

% Constriction: consider the properties of the candidate label.

Typically, a constriction can be introduced in the edge relaxation,
where we drop a candidate label if it does not meet some conditions.
For instance, to limit the length of a shortest path, we drop a
candidate label if its cost is greater than a given value.  We use the
decision function introduced in the problem statement for the
constriction.

\subsection{Algorithm}

% The typical structure.

Algorithm \ref{a:algorithm} presents the complete algorithm with the
typical Dijkstra algorithm structure, where the main loop processes
the labels of the priority queue $Q$ sorted in the ascending order of
the label cost.

\begin{algorithm}[t]
  \caption{Generic Dijkstra\\
    In: graph $G$, source vertex $s$, target vertex $t$\\
    Out: a pair of a shortest path, and a CU\\
    \emph{Here we concentrate on permanent labels $l$.}}
  \label{a:algorithm}
    \begin{algorithmic}
      \STATE $Q_s = \{\lbl{0}{\Omega}{\nulledge}\}$
      \WHILE{$Q$ is not empty}
      \STATE $l = \text{pop}(Q)$
      \STATE $e = \edge(l)$
      \STATE $v = \target(e)$
      \STATE // Add $l$ to the set of permanent labels of vertex $v$.
      \STATE $L_v = L_v \cup \{l\}$
      \IF{$v == t$}
      \STATE break the main loop
      \ENDIF
      \FORALL{out edge $e'$ of $v$ in $G$}
      \STATE $\myrelax(e', l)$
      \ENDFOR
      \ENDWHILE
      \RETURN $\text{trace}(L, t)$
    \end{algorithmic}
\end{algorithm}

% Q & L.

The priority queue $Q$ is a set of $Q_v$, i.e., $Q = \{Q_v\}$, where
$Q_v$ is the set of tentative incomparable labels of vertex $v$.  The
solution of the search is maintained in $L = \{L_v\}$, where $L_v$ is
the set of permanent (optimal) incomparable labels of vertex $v$.

% How the search starts.

To boot the search, we initialize $Q_s =
\{\lbl{0}{\Omega}{\nulledge}\}$ to make all units available at vertex
$s$ at cost 0.  The null edge $\nulledge$, which is not present in
graph $G$, marks the beginning of a shortest path.

% Iteration.

In every iteration of the main loop, we process label $l$ of the
lowest cost popped from queue $Q$, and along edge $e$ we visit vertex
$v$.  Function $\target(e)$ gives the target vertex of edge $e$ with
the special case of $\target(e_{\emptyset}) == s$, i.e., the source
vertex.  If $v == t$, then we found a solution and break the main
loop.  Otherwise, we try to relax each edge $e'$ leaving vertex $v$.

% Relaxation in general.

Algorithm \ref{a:relax} shows the relaxation of edge $e'$ reached with
label $l$.  We relax the edge for a set of incomparable candidate
labels $l'$, which we produce for each CU $C'$ in the set obtained by
intersecting the CU of label $l$ and the available units of edge $e'$.
The candidate labels $l'$ have the same cost $c'$ and edge $e'$, and
differ in the CU $C'$ only.  We examine label $l'$ if the decision
function $\decide(l')$ permits.

% How the relaxation works for label $l'$.

Next, if there is no permanent or tentative label of vertex $v'$
better than $l'$, we relax the edge by first discarding any tentative
label of vertex $v'$ which is worse than $l'$, and then adding $l'$ to
$Q_{v'}$.  Edge relaxation replenishes the queue with tentative
labels, and the algorithm keeps iterating until destination vertex $t$
is reached, or the queue is empty.

\begin{algorithm}[t]
  \caption{$\myrelax$\\
    In: edge $e'$, label $l$\\
    \emph{Here we concentrate on tentative labels $l'$.}}
  \label{a:relax}
    \begin{algorithmic}
      \STATE $c = \cost(l)$
      \STATE $c' = c + \text{cost}(e')$
      \STATE $C = \CU(l)$
      \STATE $v' = \text{target}(e')$
      \FORALL{CU $C'$ in $C \cap \AU(e')$}
      \STATE $l' = (c', C', e')$ 
      \IF{$\decide(l')$}
      \IF{$\nexists l_{v'} \in L_{v'} : l_{v'} < l'$}
      \IF{$\nexists q_{v'} \in Q_{v'} : q_{v'} < l'$}
      \STATE // Discard tentative labels $q_{v'}$ such that $l' < q_{v'}$.
      \STATE $Q_{v'} = Q_{v'} \setminus \{q_{v'} \in Q_{v'}: l' < q_{v'}\}$
      \STATE // Add $l'$ to the set of tentative labels of vertex $v'$.
      \STATE $Q_{v'} = Q_{v'} \cup \{l'\}$
      \ENDIF
      \ENDIF
      \ENDIF
      \ENDFOR
    \end{algorithmic}
\end{algorithm}

% How to take into account the spectrum allocation policy.

The required spectrum allocation policy (e.g., first-fit, best-fit)
can be taken into account by the priority queue while popping a label.
From among the labels of the lowest (equal) cost, the priority queue
should pop the label with the CU preferred by the given spectrum
allocation policy.  For instance, for the best-fit spectrum allocation
policy, the queue should pop the label not only of the lowest cost,
but also of the CU with the lowest number of units.

% The trace function.

Finally, once we leave the main loop, function $\trace(L, t)$ traces
back from node $t$ a shortest path found, if any, based on the vertex
labels $L$, and returns a pair of a path and a CU.  The function
selects, according to the spectrum allocation policy used, the CU with
the required number of units from the CU of the vertex label of node
$t$, which may have more units than required.  We do not present the
function, since it is the typical Dijkstra path back-tracing.

% This is the end: undirected multigraphs too.

To be general, we stated and solved the problem for a directed
multigraph, but the algorithm can be used for undirected multigraphs
too.  In our simulations we modeled an EON with an undirected graph,
and were able to use our algorithm.

%%%%%%%%%%%%%%%%%%%%%%%%%%%%%%%%%%%%%%%%%%%%%%%%%%%%%%%%%%%%%%%%%%%%%%%%%%%

\section{Simulations}
\label{simulations}

% What's the deal with the simulations.

We compared the memory and time performance of the proposed algorithm
applied to the dynamic RMSA problem with three other optimal
algorithms: the filtered-graphs algorithm, the brute-force algorithm,
and the unlimited Yen KSP algorithm.

% Yen is out of the gate first.

Early in our simulation studies, we realized that the unlimited Yen
KSP is very time inefficient, which prohibited using it in our
large-scale simulations.  In one case, Yen produced over three hundred
thousand shortest paths in 24 hours which had not met the spectrum
continuity and contiguity constraints.  For this reason, we were
unable to include the unlimited Yen KSP algorithm in the comparison.

% Spectrum allocation policy.

We report the simulation results only for the \emph{first-fit}
spectrum allocation policy, since the \emph{best-fit} and
\emph{random-fit} policies performed worse for all algorithms
compared.

% The comparison was fair - the same Dijkstra implementation.

To make the comparison unbiased, we implemented all algorithms with a
great attention to detail, and an emphasis on time and memory
performance.  We especially carefully treated the filtered-graphs
algorithm, and implemented it using our generic Dijkstra
implementation, which is ultra efficient, employing the latest C++17
functionality, such as the extraction of the associative-array
elements with the move semantics.

% We were careful about the algorithm and the implementation.

We made sure that the proposed algorithm and implementation were
correct with unit tests, assertions, and extra code that validated the
optimality and integrity of the results found.  We validated, with the
filtered-graphs algorithm, not only the final results found, but also
the intermediate results.  In the production runs, we disabled the
assertions and the extra code, so that the time measurements were not
disturbed.

% What we studied: LH75, but also LH25 and UD100 a bit.

In the simulations, we concentrated on the long-haul networks with 75
nodes, since they model the current and future optical transport
networks well.  However, we also carried out simulations for legacy
long-haul networks with 25 nodes, and found that the brute-force
algorithm, and the proposed algorithm performed comparably in terms of
memory usage (using tens of kB), and execution time (taking
milliseconds), while the filtered-graph algorithm was about a hundred
times slower, but used only hundreds of bytes.

% Ultra-dense networks with 100 nodes.

We also carried out fragmentary simulations for large, ultra-dense
random networks with 100 nodes and 1000 edges.  We found that the
brute-force algorithm needed very large amounts of memory (more than
96 GB), which we did not have.  The proposed algorithm on average was
using a few hundred kB, and taking milliseconds.  The filtered-graph
algorithm was on average a thousand times slower, and used only a few
kB.

\subsection{Simulation setting}
\label{setting}

The simulation setting had three major parts: the network model, the
traffic model, and the signal modulation model.

\subsubsection{Network model}

% On random graphs and traffic.

We generated a set of random long-haul graphs with random traffic to
obtain reliable statistical results for various populations of
interest.  Specifically, we used Gabriel graphs, because they have
been shown to model the properties of the long-haul transport networks
very well \cite{10.1109/ICUMT.2013.6798402}.

% On the Gabriel graphs we generate.

A network model is defined by a network graph, and the number
$|\Omega|$ of edge units.  We randomly generated 100 Gabriel graphs.
Each graph had 75 vertices, which were uniformly distributed over a
square area with the typical density of one vertex per $10^4$
$\text{km}^2$.  In generating Gabriel graphs, the number of edges
cannot be directly controlled, as it depends on the location of
vertices, and on the candidate edges meeting the conditions of the
Gabriel graph.  The statistics of the generated graphs are given in
Table \ref{t:netstats}.

For $|\Omega|$, we used the three values of 160, 320, and 640.  For
the conventional band (C-band), 160 units would require the spacing of
25 GHz, 320 units the spacing of 12.5 GHz, and 640 units the spacing
of 6.25 GHz.

\begin{table}
  \caption{Statistics of the generated Gabriel networks.}
  \label{t:netstats}
  \centering
  \begin{tabular}{|l|r|r|r|r|}
    \hline
    \textbf{Measured quantity} & \textbf{Min} & \textbf{Average} & \textbf{Max} & \textbf{Variance}\\
    \hline
    number of edges & 119 & 131.53 & 145 & 33.84\\
    edge length & 1 & 97.12 & 499 & 2712.19\\
    vertex degree & 1 & 3.51 & 8 & 1.21\\
    shortest-path hops & 1 & 5.96 & 20 & 8.62\\
    shortest-path length & 1 & 510.45 & 1369 & 59518.11\\
    \hline
  \end{tabular}
\end{table}

\subsubsection{Traffic model}

% On traffic in general.

Demands arrive according to the exponential distribution with the rate
of $\lambda$ demands per day.  The probability distribution of the
demand holding time is also exponential with the mean of $\delta$
days.  The end nodes of a demand are different and chosen at random.
The number of units a demand requests follows the distribution of
$(\Poisson(\gamma - 1) + 1)$ with the mean of $\gamma$, i.e., the
Poisson distribution shifted by one to the right, to ensure that the
number of units is greater than zero.

% The requested number of units: explanations.

We describe a demand with a requested number of units, and not with a
bitrate, to keep the discussion simple, and because the algorithms
operate on units, and not on bitrate.  If needed, a demand can be
described with bitrate, and the required number of units can be
obtained using function $n_1(b)$, which should take into account the
technical details of the specific modulation and optical hardware
used.

% Why these values of gamma.

To investigate the difference in algorithm performance as $\gamma$
increases, we used two values of 1 and 10 for $\gamma$.  Using
$\gamma = 1$ approximates the algorithm performance for a traditional
WDM network.

% Traffic is irrelevant.

We argue that the choice of a traffic model is irrelevant to our study
as the traffic only produces the input data (i.e., the state of the
graph) for the routing algorithms, and we chose the exponential and
Poisson distributions to keep the discussion simple.  The question is
how the algorithms perform under the given utilization, regardless of
how the utilization was obtained, which could have been equally well
produced randomly.

% Lambda as a function of the offered load.

We express the mean demand arrival rate $\lambda$ as a function of the
offered load $\mu$ as estimated by (\ref{e:lambda}), where $\alpha$ is
the mean number of edges of all shortest-paths in a network being
simulated.  We define the offered load $\mu$ as the ratio of the
number of units demanded to the number of units in the network.  The
average number of demanded units is $\lambda\delta\alpha\gamma$, since
there are $\lambda\delta$ active connections (assuming every demand
has a connection established), and since in an unloaded network a
connection takes on average $\alpha\gamma$ units.  The number of units
in the network is $|E||\Omega|$, and so the offered load $\mu =
\lambda\delta\alpha\gamma/(|E||\Omega|)$, from which (\ref{e:lambda})
follows.

\begin{equation}
  \lambda (\mu) = \frac{\mu|E||\Omega|}{\delta\alpha\gamma}
  \label{e:lambda}
\end{equation}

% The network utilization.

We define the \emph{network utilization} as the ratio of the number of
units in use to the total number of units on all edges.  We cannot
directly control the network utilization, but only measure it in
response to the offered load $\mu$.

\subsubsection{Signal modulation model}

% The signal modulation in the EON.

In EONs, signal modulation can be adapted to the quality of the
optical path, which depends on the length of the path and the optical
components traversed.  If we assume that the quality of the optical
path depends mostly on its length, then a modulation can be
characterized by the reach, i.e., the maximum length of a path above
which the modulation cannot be used, because the signal would suffer
unacceptable bit error rate.  The reach increases as the spectral
efficiency of the modulation decreases.

% How the modulation reach can be doubled.

In \cite{10.1109/MCOM.2010.5534599}, the authors experimentally
demonstrated that if, for a demand requesting bitrate $b$ in b/s, the
most spectrally-efficient modulation available of reach $r_M$ requires
$n_M(b)$ units, then a less spectrally-efficient modulation of reach
$r_m$ requires $(M + 1 - m)n_M(b)$ units, where $m = M, (M - 1), ...,
1$ is integer and is called the modulation level, and $M$ is the
modulation level of the most spectrally-efficient modulation
considered.  Reach $r_m$ doubles for the next less
spectrally-efficient modulation (i.e., $m$ decreases), as given by
(\ref{e:r_m}).  Therefore, for a path of length $r_M < d \le r_1$, we
need to use modulation level $m$ given by (\ref{e:m}), derived from
(\ref{e:r_m}) with the assumption that $m$ is integer.  For $d \le
r_M$ we use $m = M$, and for $r_1 < d$ we have no modulation
available.

\begin{equation}
  r_m = r_M2^{M - m}
  \label{e:r_m}
\end{equation}

\begin{equation}
  m(d) = M + 1 - \ceil{log_2(2d / r_M)}
  \label{e:m}
\end{equation}

% Too strict: m is integer.  Our model.

However, the assumption that the number of required units is an
integer multiple of $n_M(b)$, because $m$ is integer, is too strict.
The bit error rate, which is increasing with the increasing path
length $d$, can be lowered by using more units for the overhead of the
error correction codes.  In the most general case, the number of
required units should increase by one.  For this reason, we allow the
number of required units to be any integer from $n_M(b)$ to $Mn_M(b)$
depending on distance $0 \le d \le r_1$, as given by (\ref{e:n}).

% The number of required units.

\begin{equation}
  n(b, d) =
  \begin{cases}
    n_M(b) & \text{if}\ d \le r_M \\
    \infty & \text{if}\ r_1 < d \\
    \ceil{n_M(b)log_2(2d / r_M)} & \text{otherwise}%
  \end{cases}
  \label{e:n}
\end{equation}

% It's here, so that it appears at the rigth place.

\begin{algorithm}[t]
  \caption{$\decide$\\
    In: label $l'$\\
    Out: boolean}
  \label{a:decide}
  \begin{algorithmic}
    \STATE // Make sure that $l'$ has the required number of units.
    \RETURN $n(b, \cost(l')) \le |\CU(l')|$
    \end{algorithmic}
\end{algorithm}

% The decision function and the modulation.

The decision function in Algorithm \ref{a:decide} uses (\ref{e:n}) to
check whether candidate label $l'$ is able to support the demand of
bitrate $b$, i.e., whether $l'$ has at least the number of contiguous
units required for bitrate $b$ at the cost (distance) of $l'$.

% On $r_M$.

In our simulations we assumed $r_1$ equals the length of the longest
of all shortest paths multiplied by 1.5, which allows us to consider
paths which were far longer than an average shortest path.  Using
(\ref{e:r_m}), we calculated $r_M$ for (\ref{e:n}).  We tried to
increase the multiple from 1.5 to 2.0, but the brute-force algorithm
would cause the simulations to run out of memory.  We assumed $M = 4$.

\subsection{Runs and populations}

% What a simulation run is.

\emph{A simulation run} simulated 100 days of a network in operation.
The parameters of a simulation run were: the network graph, the number
of units $|\Omega|$, the mean number of demanded units $\gamma$, the
offered load $\mu$, and the mean connection holding time $\delta$.  A
simulation run reported the mean network utilization, and, for each of
the algorithms evaluated, the mean and maximum memory used, and the
mean and maximum times taken by a shortest-path search.

% How the comparison works.

When a demand arrived, we searched for an optical path using all
evaluated algorithms.  We made sure that either all algorithms found
no result, or that all results found were of the same cost and the
same number of contiguous units.  The result of the proposed algorithm
was used to establish a connection.

% Memory measurement.

During a shortest-path search we measured the maximum number of 32 bit
words required by the largest data structures of the algorithm
evaluated: the sets of tentative and permanent labels of the proposed
algorithm, the vertex labels and the priority queue of the
filtered-graphs algorithm, and the priority queue of the brute-force
algorithm.  We tracked separately the words required to store the
costs, edges, and units.  We stored a cost in one word, an edge in two
words, a single unit in one word, and a CU in two words.  Using the
obtained memory measurements of a shortest-path search, we calculated
the mean and maximum memory used by each algorithm throughout a
simulation run.  With careful implementation and testing, we made sure
that the memory measurement had negligible effect on the time
measurement.

% Time measurement problematic.

Time measurement required special attention, because we ran
simulations using a supercomputing infrastructure.  While we were able
to select one specific hardware type for all our simulations, we had
little control over how much the hardware was loaded with the
processes of other users, which could have severely degraded the
performance of our simulations.  For this reason we repeated a
simulation run after a few hours, and for every time measurement, we
took the minimum of the two values obtained, based on which we
calculated the mean and maximum time taken by each of the algorithms
throughout a simulation run.

% The populations, and the general idea.

We are interested in the results for a statistical population of
simulation runs, rather than in the results for a single simulation
run only, because a simulation run could have been an outlier with
unusual results due to its randomly-generated network and traffic.  To
estimate a mean result for a population, we carried out the simulation
runs which were the population samples, and calculated a \emph{sample
mean} of all the mean results reported by the simulation runs.  We
estimated the pessimistic algorithm performance of a population by a
\emph{sample maximum}, which is the maximum of the maximum results
reported by the simulation runs.

% The parameters of a population, and the details.

In a given population there were 100 simulation runs whose parameters
differed only with the network model.  Hence, we had 100 Gabriel
graphs generated, and used for every population.  We had 102
populations, because we varied 3 values of the number of units
$|\Omega|$ (160, 320, 640), 17 values of the offered load $\mu$ (0.05,
0.075, 0.1, 0.125, 0.15, 0.175, 0.2, 0.3, 0.45, 0.55, 0.65, 0.75, 1,
1.25, 1.5, 1.75, and 2), and two values of the mean number of demanded
units $\gamma$ (1, 10).  For all populations, the mean connection
holding time $\delta = 10$ days was constant.  In total we carried out
10200 simulation runs (102 populations $\times$ 100 samples), which
then we repeated.  For the proposed algorithm and the filtered-graphs
algorithms, the sample means credibly estimate the mean results of
populations, since their relative standard error is usually around
1\%.  For the brute-force algorithm, the sample means frequently have
the relative standard error of around 20\%.

%%%%%%%%%%%%%%%%%%%%%%%%%%%%%%%%%%%%%%%%%%%%%%%%%%%%%%%%%%%%%%%%%%%%%%%%%%%

\subsection{Simulation results}
\label{results}

% The curves in the plots.

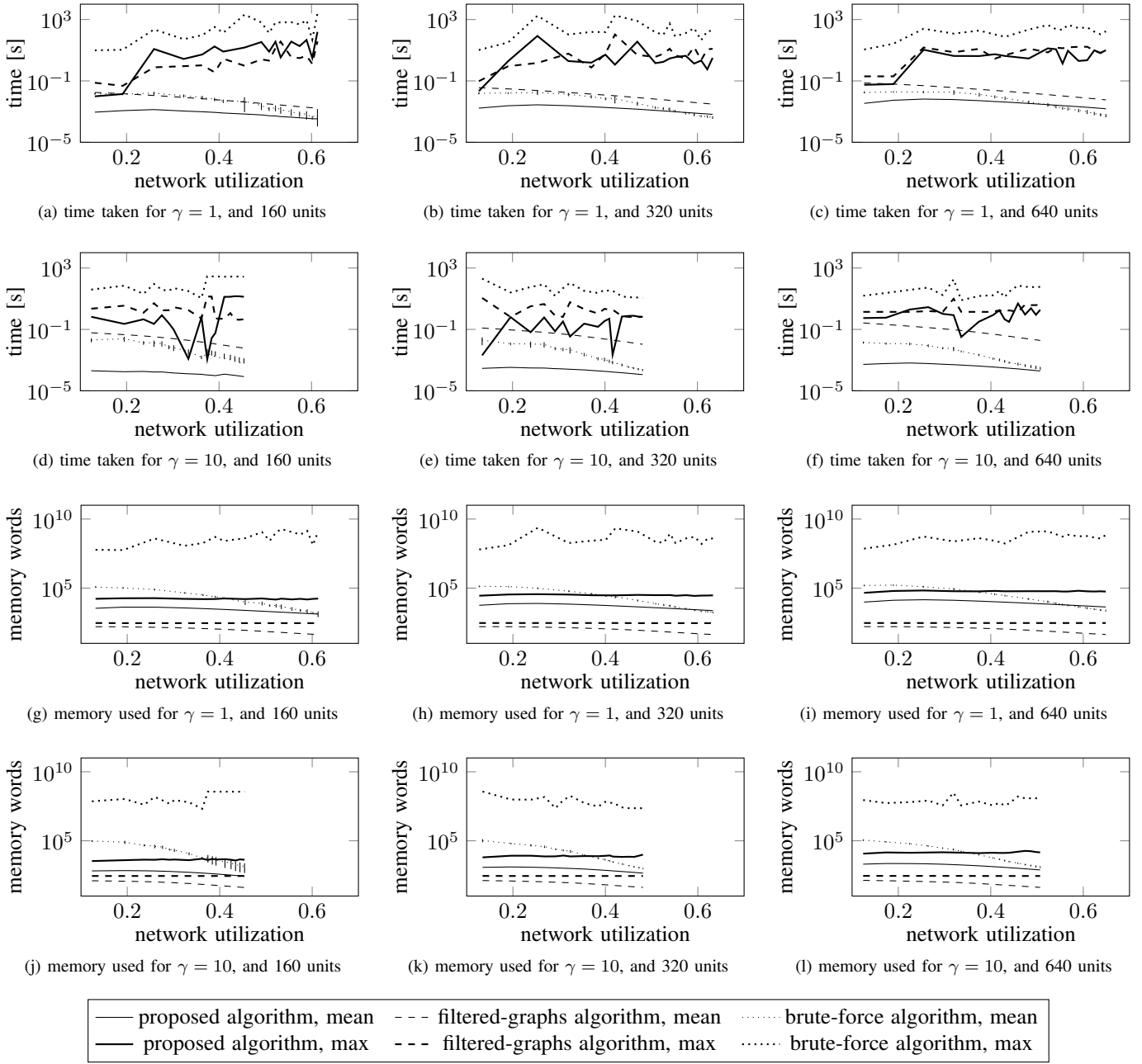
\begin{figure*}
  \subfloat[time taken for $\gamma = 1$, and 160 units]{%
    \label{f:time_1_160}%
    \begin{tikzpicture}
\begin{semilogyaxis}[xlabel = {network utilization}, ylabel = {time [s]}, xmin = 0.1, xmax = 0.7, height = 3.8 cm, width = 6 cm, xlabel shift = -3 pt, legend columns = 3, legend to name = regular, legend style = {/tikz/every even column/.append style = {column sep = 0.25 cm}}, ymin = 1e-5, ymax = 1e4, ylabel shift = -6 pt]
\addplot[solid]
coordinates {
(0.131461, 0.000915531) +- (0.00071791, 2.50967e-05)
(0.191483, 0.00115415) +- (0.00112687, 3.31355e-05)
(0.259328, 0.00134909) +- (0.00218623, 0.000139242)
(0.323594, 0.00108787) +- (0.00252102, 6.75022e-05)
(0.363878, 0.000972387) +- (0.00283414, 6.84132e-05)
(0.390379, 0.000889907) +- (0.00337979, 6.89787e-05)
(0.409166, 0.000806853) +- (0.00376252, 3.28894e-05)
(0.4551, 0.000693067) +- (0.0042842, 3.94662e-05)
(0.491548, 0.000606982) +- (0.00443894, 5.76888e-05)
(0.508246, 0.000535949) +- (0.00465782, 1.90037e-05)
(0.522032, 0.000518915) +- (0.00459475, 3.45546e-05)
(0.533248, 0.000471594) +- (0.00447518, 1.17386e-05)
(0.556131, 0.000430615) +- (0.00432393, 2.07042e-05)
(0.574202, 0.00039355) +- (0.00414095, 1.15789e-05)
(0.589254, 0.000369359) +- (0.00398702, 1.79954e-05)
(0.60164, 0.000340327) +- (0.0040399, 6.83603e-06)
(0.612831, 0.000338187) +- (0.00389073, 4.01443e-05)
};
\addlegendentry{proposed algorithm, mean}
\addplot[dashed]
coordinates {
(0.131461, 0.0177492) +- (0.00071791, 0.000322044)
(0.191483, 0.0146806) +- (0.00112687, 0.000330613)
(0.259328, 0.0112019) +- (0.00218623, 0.000310254)
(0.323594, 0.00838239) +- (0.00252102, 0.000212618)
(0.363878, 0.00689054) +- (0.00283414, 0.000135688)
(0.390379, 0.00603422) +- (0.00337979, 0.000106688)
(0.409166, 0.00546432) +- (0.00376252, 8.64941e-05)
(0.4551, 0.00422874) +- (0.0042842, 6.77692e-05)
(0.491548, 0.00342434) +- (0.00443894, 5.58205e-05)
(0.508246, 0.00310789) +- (0.00465782, 5.06793e-05)
(0.522032, 0.00286176) +- (0.00459475, 4.86398e-05)
(0.533248, 0.00268172) +- (0.00447518, 5.04706e-05)
(0.556131, 0.00231951) +- (0.00432393, 3.80442e-05)
(0.574202, 0.00208305) +- (0.00414095, 3.26035e-05)
(0.589254, 0.00190356) +- (0.00398702, 2.80407e-05)
(0.60164, 0.00176021) +- (0.0040399, 2.63544e-05)
(0.612831, 0.00164054) +- (0.00389073, 2.55555e-05)
};
\addlegendentry{filtered-graphs algorithm, mean}
\addplot[dotted, mybars]
coordinates {
(0.131461, 0.0146866) +- (0.00071791, 0.00212556)
(0.191483, 0.0136251) +- (0.00112687, 0.0019204)
(0.259328, 0.0159283) +- (0.00218623, 0.00360882)
(0.323594, 0.0106784) +- (0.00252102, 0.00232435)
(0.363878, 0.00836192) +- (0.00283414, 0.00212537)
(0.390379, 0.0063022) +- (0.00337979, 0.00157151)
(0.409166, 0.00484191) +- (0.00376252, 0.00119507)
(0.4551, 0.00478716) +- (0.0042842, 0.00386298)
(0.491548, 0.00244181) +- (0.00443894, 0.00116373)
(0.508246, 0.00150375) +- (0.00465782, 0.00040853)
(0.522032, 0.00176898) +- (0.00459475, 0.000730174)
(0.533248, 0.00136585) +- (0.00447518, 0.000642871)
(0.556131, 0.000968476) +- (0.00432393, 0.000448859)
(0.574202, 0.00071866) +- (0.00414095, 0.000269328)
(0.589254, 0.000674063) +- (0.00398702, 0.00037399)
(0.60164, 0.000410082) +- (0.0040399, 9.21101e-05)
(0.612831, 0.000787444) +- (0.00389073, 0.000679186)
};
\addlegendentry{brute-force algorithm, mean}
\addplot[thick, solid]
coordinates {
(0.131461, 0.00986607) +- (0.00071791, 0)
(0.191483, 0.0139366) +- (0.00112687, 0)
(0.259328, 11.9558) +- (0.00218623, 0)
(0.323594, 2.69241) +- (0.00252102, 0)
(0.363878, 5.27588) +- (0.00283414, 0)
(0.390379, 17.549) +- (0.00337979, 0)
(0.409166, 8.12451) +- (0.00376252, 0)
(0.4551, 14.9258) +- (0.0042842, 0)
(0.491548, 34.1251) +- (0.00443894, 0)
(0.508246, 8.05195) +- (0.00465782, 0)
(0.522032, 35.7323) +- (0.00459475, 0)
(0.533248, 3.40165) +- (0.00447518, 0)
(0.556131, 36.4807) +- (0.00432393, 0)
(0.574202, 18.5178) +- (0.00414095, 0)
(0.589254, 46.6881) +- (0.00398702, 0)
(0.60164, 3.13923) +- (0.0040399, 0)
(0.612831, 156.138) +- (0.00389073, 0)
};
\addlegendentry{proposed algorithm, max}
\addplot[thick, dashed]
coordinates {
(0.131461, 0.0755364) +- (0.00071791, 0)
(0.191483, 0.0475364) +- (0.00112687, 0)
(0.259328, 0.767084) +- (0.00218623, 0)
(0.323594, 0.910752) +- (0.00252102, 0)
(0.363878, 1.03613) +- (0.00283414, 0)
(0.390379, 0.36614) +- (0.00337979, 0)
(0.409166, 2.79573) +- (0.00376252, 0)
(0.4551, 0.9727) +- (0.0042842, 0)
(0.491548, 2.53161) +- (0.00443894, 0)
(0.508246, 2.095) +- (0.00465782, 0)
(0.522032, 13.7297) +- (0.00459475, 0)
(0.533248, 36.2483) +- (0.00447518, 0)
(0.556131, 3.5139) +- (0.00432393, 0)
(0.574202, 3.06223) +- (0.00414095, 0)
(0.589254, 4.30653) +- (0.00398702, 0)
(0.60164, 1.29736) +- (0.0040399, 0)
(0.612831, 37.5414) +- (0.00389073, 0)
};
\addlegendentry{filtered-graphs algorithm, max}
\addplot[thick, dotted]
coordinates {
(0.131461, 9.79568) +- (0.00071791, 0)
(0.191483, 10.6847) +- (0.00112687, 0)
(0.259328, 225.454) +- (0.00218623, 0)
(0.323594, 52.1382) +- (0.00252102, 0)
(0.363878, 96.379) +- (0.00283414, 0)
(0.390379, 327.462) +- (0.00337979, 0)
(0.409166, 137.214) +- (0.00376252, 0)
(0.4551, 2086.92) +- (0.0042842, 0)
(0.491548, 647.765) +- (0.00443894, 0)
(0.508246, 132.371) +- (0.00465782, 0)
(0.522032, 648.058) +- (0.00459475, 0)
(0.533248, 646.211) +- (0.00447518, 0)
(0.556131, 668.84) +- (0.00432393, 0)
(0.574202, 480.789) +- (0.00414095, 0)
(0.589254, 898.713) +- (0.00398702, 0)
(0.60164, 49.1793) +- (0.0040399, 0)
(0.612831, 2481.34) +- (0.00389073, 0)
};
\addlegendentry{brute-force algorithm, max}
\end{semilogyaxis}
\end{tikzpicture}}\hfill%
  \subfloat[time taken for $\gamma = 1$, and 320 units]{%
    \label{f:time_1_320}%
    \begin{tikzpicture}
\begin{semilogyaxis}[xlabel = {network utilization}, ylabel = {time [s]}, xmin = 0.1, xmax = 0.7, height = 3.8 cm, width = 6 cm, xlabel shift = -3 pt, legend columns = 3, legend to name = regular, legend style = {/tikz/every even column/.append style = {column sep = 0.25 cm}}, ymin = 1e-5, ymax = 1e4, ylabel shift = -6 pt]
\addplot[solid]
coordinates {
(0.127142, 0.00167973) +- (0.000588861, 5.27242e-05)
(0.19061, 0.00240826) +- (0.000927054, 7.64082e-05)
(0.253873, 0.00276366) +- (0.00161753, 0.000241344)
(0.320806, 0.00241569) +- (0.00220855, 9.02573e-05)
(0.370566, 0.002058) +- (0.00265177, 6.49092e-05)
(0.401078, 0.00185603) +- (0.00336707, 5.59983e-05)
(0.421552, 0.00171521) +- (0.00382383, 4.71634e-05)
(0.469919, 0.00143768) +- (0.00453692, 5.14703e-05)
(0.50895, 0.00119569) +- (0.00480454, 3.27814e-05)
(0.526206, 0.001108) +- (0.00483655, 2.96503e-05)
(0.53985, 0.00103877) +- (0.00486905, 2.75275e-05)
(0.551682, 0.000981867) +- (0.00476024, 2.528e-05)
(0.574532, 0.000874972) +- (0.00456749, 2.12327e-05)
(0.59286, 0.000803498) +- (0.00435369, 1.87135e-05)
(0.608082, 0.000745184) +- (0.00420392, 1.69564e-05)
(0.621302, 0.000699682) +- (0.00412613, 1.54777e-05)
(0.632771, 0.000660365) +- (0.00406718, 1.47367e-05)
};
\addlegendentry{proposed algorithm, mean}
\addplot[dashed]
coordinates {
(0.127142, 0.0360233) +- (0.000588861, 0.000617802)
(0.19061, 0.0301694) +- (0.000927054, 0.000653938)
(0.253873, 0.0236477) +- (0.00161753, 0.000636873)
(0.320806, 0.0175467) +- (0.00220855, 0.000490874)
(0.370566, 0.0138274) +- (0.00265177, 0.000301235)
(0.401078, 0.0118667) +- (0.00336707, 0.00022449)
(0.421552, 0.0107787) +- (0.00382383, 0.000271172)
(0.469919, 0.0081982) +- (0.00453692, 0.000153352)
(0.50895, 0.00652696) +- (0.00480454, 0.0001242)
(0.526206, 0.00589577) +- (0.00483655, 0.000113651)
(0.53985, 0.00541144) +- (0.00486905, 0.0001037)
(0.551682, 0.00504531) +- (0.00476024, 9.47204e-05)
(0.574532, 0.00436947) +- (0.00456749, 7.89776e-05)
(0.59286, 0.00392269) +- (0.00435369, 6.80254e-05)
(0.608082, 0.00357218) +- (0.00420392, 6.27182e-05)
(0.621302, 0.00329915) +- (0.00412613, 5.56088e-05)
(0.632771, 0.00306763) +- (0.00406718, 4.97775e-05)
};
\addlegendentry{filtered-graphs algorithm, mean}
\addplot[dotted, mybars]
coordinates {
(0.127142, 0.0156475) +- (0.000588861, 0.00210153)
(0.19061, 0.0161574) +- (0.000927054, 0.00216846)
(0.253873, 0.017429) +- (0.00161753, 0.00545774)
(0.320806, 0.0133309) +- (0.00220855, 0.00318763)
(0.370566, 0.00937222) +- (0.00265177, 0.00220924)
(0.401078, 0.00687968) +- (0.00336707, 0.00133998)
(0.421552, 0.00693868) +- (0.00382383, 0.00353921)
(0.469919, 0.00337923) +- (0.00453692, 0.000954309)
(0.50895, 0.00184657) +- (0.00480454, 0.000348168)
(0.526206, 0.00157415) +- (0.00483655, 0.000309076)
(0.53985, 0.00143162) +- (0.00486905, 0.000401974)
(0.551682, 0.00107566) +- (0.00476024, 0.000185961)
(0.574532, 0.000793511) +- (0.00456749, 0.000134286)
(0.59286, 0.000646035) +- (0.00435369, 0.000146869)
(0.608082, 0.000507057) +- (0.00420392, 7.53285e-05)
(0.621302, 0.00047307) +- (0.00412613, 8.551e-05)
(0.632771, 0.000414937) +- (0.00406718, 8.04314e-05)
};
\addlegendentry{brute-force algorithm, mean}
\addplot[thick, solid]
coordinates {
(0.127142, 0.0215171) +- (0.000588861, 0)
(0.19061, 1.94419) +- (0.000927054, 0)
(0.253873, 84.8905) +- (0.00161753, 0)
(0.320806, 1.96279) +- (0.00220855, 0)
(0.370566, 1.54636) +- (0.00265177, 0)
(0.401078, 5.06946) +- (0.00336707, 0)
(0.421552, 1.1404) +- (0.00382383, 0)
(0.469919, 36.0741) +- (0.00453692, 0)
(0.50895, 1.4369) +- (0.00480454, 0)
(0.526206, 1.73415) +- (0.00483655, 0)
(0.53985, 2.87924) +- (0.00486905, 0)
(0.551682, 3.52297) +- (0.00476024, 0)
(0.574532, 4.32567) +- (0.00456749, 0)
(0.59286, 1.34873) +- (0.00435369, 0)
(0.608082, 2.13068) +- (0.00420392, 0)
(0.621302, 0.566306) +- (0.00412613, 0)
(0.632771, 3.16) +- (0.00406718, 0)
};
\addlegendentry{proposed algorithm, max}
\addplot[thick, dashed]
coordinates {
(0.127142, 0.0955431) +- (0.000588861, 0)
(0.19061, 0.904737) +- (0.000927054, 0)
(0.253873, 1.47755) +- (0.00161753, 0)
(0.320806, 5.98589) +- (0.00220855, 0)
(0.370566, 0.74337) +- (0.00265177, 0)
(0.401078, 6.07238) +- (0.00336707, 0)
(0.421552, 105.38) +- (0.00382383, 0)
(0.469919, 4.87848) +- (0.00453692, 0)
(0.50895, 3.67276) +- (0.00480454, 0)
(0.526206, 5.87116) +- (0.00483655, 0)
(0.53985, 12.468) +- (0.00486905, 0)
(0.551682, 4.72929) +- (0.00476024, 0)
(0.574532, 2.84653) +- (0.00456749, 0)
(0.59286, 16.1147) +- (0.00435369, 0)
(0.608082, 2.26139) +- (0.00420392, 0)
(0.621302, 11.368) +- (0.00412613, 0)
(0.632771, 12.3556) +- (0.00406718, 0)
};
\addlegendentry{filtered-graphs algorithm, max}
\addplot[thick, dotted]
coordinates {
(0.127142, 10.2823) +- (0.000588861, 0)
(0.19061, 32.2074) +- (0.000927054, 0)
(0.253873, 1764.25) +- (0.00161753, 0)
(0.320806, 106.113) +- (0.00220855, 0)
(0.370566, 155.01) +- (0.00265177, 0)
(0.401078, 152.739) +- (0.00336707, 0)
(0.421552, 1842.29) +- (0.00382383, 0)
(0.469919, 809.652) +- (0.00453692, 0)
(0.50895, 147.619) +- (0.00480454, 0)
(0.526206, 100.603) +- (0.00483655, 0)
(0.53985, 552.546) +- (0.00486905, 0)
(0.551682, 80.1287) +- (0.00476024, 0)
(0.574532, 130.819) +- (0.00456749, 0)
(0.59286, 274.925) +- (0.00435369, 0)
(0.608082, 44.939) +- (0.00420392, 0)
(0.621302, 173.957) +- (0.00412613, 0)
(0.632771, 185.418) +- (0.00406718, 0)
};
\addlegendentry{brute-force algorithm, max}
\end{semilogyaxis}
\end{tikzpicture}}\hfill%
  \subfloat[time taken for $\gamma = 1$, and 640 units]{%
    \label{f:time_1_640}%
    \begin{tikzpicture}
\begin{semilogyaxis}[xlabel = {network utilization}, ylabel = {time [s]}, xmin = 0.1, xmax = 0.7, height = 3.8 cm, width = 6 cm, xlabel shift = -3 pt, legend columns = 3, legend to name = regular, legend style = {/tikz/every even column/.append style = {column sep = 0.25 cm}}, ymin = 1e-5, ymax = 1e4, ylabel shift = -6 pt]
\addplot[solid]
coordinates {
(0.125972, 0.00344226) +- (0.000588153, 0.00011778)
(0.189211, 0.00548279) +- (0.000881459, 0.000203508)
(0.254461, 0.00654521) +- (0.00136103, 0.000245622)
(0.319807, 0.00608587) +- (0.00209886, 0.000226293)
(0.3748, 0.00516208) +- (0.00262717, 0.000182958)
(0.40838, 0.00452458) +- (0.00343446, 0.00014693)
(0.430671, 0.00419825) +- (0.00395443, 0.000146304)
(0.481924, 0.0034041) +- (0.00470082, 0.000114755)
(0.52315, 0.00282631) +- (0.00497199, 8.87601e-05)
(0.540775, 0.0026082) +- (0.00512155, 8.33976e-05)
(0.555105, 0.00244356) +- (0.00507492, 7.66212e-05)
(0.566632, 0.00230859) +- (0.0049505, 6.81386e-05)
(0.589688, 0.00208278) +- (0.0047501, 6.13014e-05)
(0.608263, 0.00190018) +- (0.0045344, 5.38659e-05)
(0.623658, 0.0017517) +- (0.00438106, 4.62132e-05)
(0.636988, 0.00163505) +- (0.00429597, 4.24098e-05)
(0.648684, 0.00151876) +- (0.00424394, 3.75272e-05)
};
\addlegendentry{proposed algorithm, mean}
\addplot[dashed]
coordinates {
(0.125972, 0.073258) +- (0.000588153, 0.0012565)
(0.189211, 0.0622393) +- (0.000881459, 0.00133116)
(0.254461, 0.0489803) +- (0.00136103, 0.00129569)
(0.319807, 0.0366091) +- (0.00209886, 0.0010483)
(0.3748, 0.0282076) +- (0.00262717, 0.000644959)
(0.40838, 0.0238191) +- (0.00343446, 0.000465265)
(0.430671, 0.0213178) +- (0.00395443, 0.000414968)
(0.481924, 0.0160605) +- (0.00470082, 0.000305706)
(0.52315, 0.0126678) +- (0.00497199, 0.00025073)
(0.540775, 0.0113812) +- (0.00512155, 0.000237707)
(0.555105, 0.0104478) +- (0.00507492, 0.000217414)
(0.566632, 0.00972322) +- (0.0049505, 0.000197315)
(0.589688, 0.00843921) +- (0.0047501, 0.000170319)
(0.608263, 0.00750785) +- (0.0045344, 0.000142425)
(0.623658, 0.00682762) +- (0.00438106, 0.000123824)
(0.636988, 0.00627333) +- (0.00429597, 0.000107078)
(0.648684, 0.00580705) +- (0.00424394, 9.8142e-05)
};
\addlegendentry{filtered-graphs algorithm, mean}
\addplot[dotted, mybars]
coordinates {
(0.125972, 0.0176937) +- (0.000588153, 0.0022744)
(0.189211, 0.0192912) +- (0.000881459, 0.00246385)
(0.254461, 0.0187659) +- (0.00136103, 0.00288704)
(0.319807, 0.0184741) +- (0.00209886, 0.00605731)
(0.3748, 0.0130431) +- (0.00262717, 0.0034546)
(0.40838, 0.00928919) +- (0.00343446, 0.00200646)
(0.430671, 0.00751822) +- (0.00395443, 0.00185026)
(0.481924, 0.00421815) +- (0.00470082, 0.000963102)
(0.52315, 0.00281739) +- (0.00497199, 0.000739828)
(0.540775, 0.00209104) +- (0.00512155, 0.000451024)
(0.555105, 0.00169931) +- (0.00507492, 0.000354474)
(0.566632, 0.00155252) +- (0.0049505, 0.000370844)
(0.589688, 0.00120666) +- (0.0047501, 0.000309453)
(0.608263, 0.000977469) +- (0.0045344, 0.000287378)
(0.623658, 0.000764271) +- (0.00438106, 0.000190021)
(0.636988, 0.000637834) +- (0.00429597, 0.000138)
(0.648684, 0.000561391) +- (0.00424394, 0.000134071)
};
\addlegendentry{brute-force algorithm, mean}
\addplot[thick, solid]
coordinates {
(0.125972, 0.0555029) +- (0.000588153, 0)
(0.189211, 0.0620519) +- (0.000881459, 0)
(0.254461, 10.7143) +- (0.00136103, 0)
(0.319807, 4.22437) +- (0.00209886, 0)
(0.3748, 4.16161) +- (0.00262717, 0)
(0.40838, 5.46905) +- (0.00343446, 0)
(0.430671, 4.22256) +- (0.00395443, 0)
(0.481924, 2.88605) +- (0.00470082, 0)
(0.52315, 13.2219) +- (0.00497199, 0)
(0.540775, 11.5591) +- (0.00512155, 0)
(0.555105, 1.34718) +- (0.00507492, 0)
(0.566632, 9.22605) +- (0.0049505, 0)
(0.589688, 2.00012) +- (0.0047501, 0)
(0.608263, 2.21893) +- (0.0045344, 0)
(0.623658, 9.82524) +- (0.00438106, 0)
(0.636988, 6.63038) +- (0.00429597, 0)
(0.648684, 9.96913) +- (0.00424394, 0)
};
\addlegendentry{proposed algorithm, max}
\addplot[thick, dashed]
coordinates {
(0.125972, 0.197324) +- (0.000588153, 0)
(0.189211, 0.198529) +- (0.000881459, 0)
(0.254461, 14.9237) +- (0.00136103, 0)
(0.319807, 7.42265) +- (0.00209886, 0)
(0.3748, 12.2751) +- (0.00262717, 0)
(0.40838, 2.79602) +- (0.00343446, 0)
(0.430671, 4.72017) +- (0.00395443, 0)
(0.481924, 12.6695) +- (0.00470082, 0)
(0.52315, 7.22824) +- (0.00497199, 0)
(0.540775, 14.5825) +- (0.00512155, 0)
(0.555105, 13.4664) +- (0.00507492, 0)
(0.566632, 15.3838) +- (0.0049505, 0)
(0.589688, 16.6979) +- (0.0047501, 0)
(0.608263, 16.759) +- (0.0045344, 0)
(0.623658, 9.4193) +- (0.00438106, 0)
(0.636988, 6.51564) +- (0.00429597, 0)
(0.648684, 10.4241) +- (0.00424394, 0)
};
\addlegendentry{filtered-graphs algorithm, max}
\addplot[thick, dotted]
coordinates {
(0.125972, 11.107) +- (0.000588153, 0)
(0.189211, 26.5255) +- (0.000881459, 0)
(0.254461, 250.778) +- (0.00136103, 0)
(0.319807, 120.357) +- (0.00209886, 0)
(0.3748, 190.641) +- (0.00262717, 0)
(0.40838, 106.159) +- (0.00343446, 0)
(0.430671, 74.4536) +- (0.00395443, 0)
(0.481924, 222.744) +- (0.00470082, 0)
(0.52315, 439.128) +- (0.00497199, 0)
(0.540775, 235.969) +- (0.00512155, 0)
(0.555105, 213.009) +- (0.00507492, 0)
(0.566632, 447.892) +- (0.0049505, 0)
(0.589688, 276.528) +- (0.0047501, 0)
(0.608263, 308.364) +- (0.0045344, 0)
(0.623658, 161.73) +- (0.00438106, 0)
(0.636988, 109.308) +- (0.00429597, 0)
(0.648684, 171.351) +- (0.00424394, 0)
};
\addlegendentry{brute-force algorithm, max}
\end{semilogyaxis}
\end{tikzpicture}}\\[3pt]
  \subfloat[time taken for $\gamma = 10$, and 160 units]{%
    \label{f:time_10_160}%
    \begin{tikzpicture}
\begin{semilogyaxis}[xlabel = {network utilization}, ylabel = {time [s]}, xmin = 0.1, xmax = 0.7, height = 3.8 cm, width = 6 cm, xlabel shift = -3 pt, legend columns = 3, legend to name = regular, legend style = {/tikz/every even column/.append style = {column sep = 0.25 cm}}, ymin = 1e-5, ymax = 1e4, ylabel shift = -6 pt]
\addplot[solid]
coordinates {
(0.123634, 0.000201211) +- (0.0017087, 7.32839e-05)
(0.194931, 0.000174658) +- (0.00286287, 1.73708e-05)
(0.236609, 0.000182214) +- (0.00297472, 2.51357e-05)
(0.259264, 0.000167571) +- (0.00314873, 1.35537e-05)
(0.276294, 0.000168126) +- (0.00342659, 2.85543e-05)
(0.291206, 0.000154036) +- (0.0031663, 8.99e-06)
(0.301845, 0.000144583) +- (0.00336988, 4.42343e-06)
(0.333706, 0.000127924) +- (0.00347518, 2.68316e-06)
(0.362026, 0.000117828) +- (0.00357163, 8.78743e-06)
(0.374476, 0.000106799) +- (0.00358651, 1.97013e-06)
(0.383973, 0.000101831) +- (0.00353857, 1.89377e-06)
(0.391935, 9.67992e-05) +- (0.00352054, 1.86941e-06)
(0.411353, 0.000121127) +- (0.00341638, 6.81755e-05)
(0.424501, 0.000108968) +- (0.00327267, 5.64841e-05)
(0.43423, 0.000100437) +- (0.00307809, 5.04194e-05)
(0.443782, 9.21006e-05) +- (0.00323497, 4.28878e-05)
(0.453735, 8.34496e-05) +- (0.00314611, 3.43028e-05)
};
\addlegendentry{proposed algorithm, mean}
\addplot[dashed]
coordinates {
(0.123634, 0.0599781) +- (0.0017087, 0.00119976)
(0.194931, 0.045453) +- (0.00286287, 0.000900031)
(0.236609, 0.037579) +- (0.00297472, 0.000703802)
(0.259264, 0.0330575) +- (0.00314873, 0.000668329)
(0.276294, 0.0294994) +- (0.00342659, 0.000547344)
(0.291206, 0.0269021) +- (0.0031663, 0.000483395)
(0.301845, 0.0247303) +- (0.00336988, 0.000436096)
(0.333706, 0.0194011) +- (0.00347518, 0.000329942)
(0.362026, 0.015046) +- (0.00357163, 0.000256214)
(0.374476, 0.0134757) +- (0.00358651, 0.000255132)
(0.383973, 0.0122459) +- (0.00353857, 0.000222506)
(0.391935, 0.0112283) +- (0.00352054, 0.000192995)
(0.411353, 0.0092714) +- (0.00341638, 0.000154564)
(0.424501, 0.00812205) +- (0.00327267, 0.000140161)
(0.43423, 0.00727965) +- (0.00307809, 0.000121695)
(0.443782, 0.00661473) +- (0.00323497, 0.000107206)
(0.453735, 0.00603316) +- (0.00314611, 9.74697e-05)
};
\addlegendentry{filtered-graphs algorithm, mean}
\addplot[dotted, mybars]
coordinates {
(0.123634, 0.0202704) +- (0.0017087, 0.00560353)
(0.194931, 0.0244099) +- (0.00286287, 0.00845911)
(0.236609, 0.0135163) +- (0.00297472, 0.00356039)
(0.259264, 0.0129446) +- (0.00314873, 0.00570287)
(0.276294, 0.0121618) +- (0.00342659, 0.00452714)
(0.291206, 0.00716413) +- (0.0031663, 0.00206066)
(0.301845, 0.00657986) +- (0.00336988, 0.00272223)
(0.333706, 0.00359233) +- (0.00347518, 0.00139321)
(0.362026, 0.00154001) +- (0.00357163, 0.00038677)
(0.374476, 0.00254678) +- (0.00358651, 0.00268074)
(0.383973, 0.00196709) +- (0.00353857, 0.00222396)
(0.391935, 0.00174164) +- (0.00352054, 0.00196719)
(0.411353, 0.00123555) +- (0.00341638, 0.00140566)
(0.424501, 0.00104294) +- (0.00327267, 0.0011649)
(0.43423, 0.000815672) +- (0.00307809, 0.000978656)
(0.443782, 0.000684368) +- (0.00323497, 0.00082212)
(0.453735, 0.000589847) +- (0.00314611, 0.000703118)
};
\addlegendentry{brute-force algorithm, mean}
\addplot[thick, solid]
coordinates {
(0.123634, 0.662487) +- (0.0017087, 0)
(0.194931, 0.228507) +- (0.00286287, 0)
(0.236609, 0.455504) +- (0.00297472, 0)
(0.259264, 0.227048) +- (0.00314873, 0)
(0.276294, 0.839567) +- (0.00342659, 0)
(0.291206, 0.257085) +- (0.0031663, 0)
(0.301845, 0.0967567) +- (0.00336988, 0)
(0.333706, 0.00121333) +- (0.00347518, 0)
(0.362026, 0.618699) +- (0.00357163, 0)
(0.374476, 0.00113476) +- (0.00358651, 0)
(0.383973, 0.0269316) +- (0.00353857, 0)
(0.391935, 0.0573126) +- (0.00352054, 0)
(0.411353, 13.1999) +- (0.00341638, 0)
(0.424501, 13.3959) +- (0.00327267, 0)
(0.43423, 14.5342) +- (0.00307809, 0)
(0.443782, 14.3524) +- (0.00323497, 0)
(0.453735, 13.3198) +- (0.00314611, 0)
};
\addlegendentry{proposed algorithm, max}
\addplot[thick, dashed]
coordinates {
(0.123634, 2.23651) +- (0.0017087, 0)
(0.194931, 3.50865) +- (0.00286287, 0)
(0.236609, 1.2261) +- (0.00297472, 0)
(0.259264, 5.0376) +- (0.00314873, 0)
(0.276294, 1.75308) +- (0.00342659, 0)
(0.291206, 1.77085) +- (0.0031663, 0)
(0.301845, 2.82879) +- (0.00336988, 0)
(0.333706, 1.80643) +- (0.00347518, 0)
(0.362026, 0.511883) +- (0.00357163, 0)
(0.374476, 13.9083) +- (0.00358651, 0)
(0.383973, 13.7544) +- (0.00353857, 0)
(0.391935, 0.516349) +- (0.00352054, 0)
(0.411353, 0.483179) +- (0.00341638, 0)
(0.424501, 1.15492) +- (0.00327267, 0)
(0.43423, 0.437768) +- (0.00307809, 0)
(0.443782, 0.425136) +- (0.00323497, 0)
(0.453735, 0.454696) +- (0.00314611, 0)
};
\addlegendentry{filtered-graphs algorithm, max}
\addplot[thick, dotted]
coordinates {
(0.123634, 39.2702) +- (0.0017087, 0)
(0.194931, 70.2881) +- (0.00286287, 0)
(0.236609, 20.6667) +- (0.00297472, 0)
(0.259264, 91.2401) +- (0.00314873, 0)
(0.276294, 37.8255) +- (0.00342659, 0)
(0.291206, 33.4026) +- (0.0031663, 0)
(0.301845, 49.0259) +- (0.00336988, 0)
(0.333706, 31.1417) +- (0.00347518, 0)
(0.362026, 10.8493) +- (0.00357163, 0)
(0.374476, 288.138) +- (0.00358651, 0)
(0.383973, 284.51) +- (0.00353857, 0)
(0.391935, 282.978) +- (0.00352054, 0)
(0.411353, 271.687) +- (0.00341638, 0)
(0.424501, 274.897) +- (0.00327267, 0)
(0.43423, 283.035) +- (0.00307809, 0)
(0.443782, 275.575) +- (0.00323497, 0)
(0.453735, 272.469) +- (0.00314611, 0)
};
\addlegendentry{brute-force algorithm, max}
\end{semilogyaxis}
\end{tikzpicture}}\hfill%
  \subfloat[time taken for $\gamma = 10$, and 320 units]{%
    \label{f:time_10_320}%
    \begin{tikzpicture}
\begin{semilogyaxis}[xlabel = {network utilization}, ylabel = {time [s]}, xmin = 0.1, xmax = 0.7, height = 3.8 cm, width = 6 cm, xlabel shift = -3 pt, legend columns = 3, legend to name = regular, legend style = {/tikz/every even column/.append style = {column sep = 0.25 cm}}, ymin = 1e-5, ymax = 1e4, ylabel shift = -6 pt]
\addplot[solid]
coordinates {
(0.134342, 0.000288456) +- (0.00132285, 7.63055e-06)
(0.195808, 0.000334802) +- (0.00155259, 2.68854e-05)
(0.239173, 0.000306479) +- (0.00200563, 8.51206e-06)
(0.26841, 0.00030268) +- (0.00204706, 1.64313e-05)
(0.292037, 0.00027647) +- (0.00259634, 7.21299e-06)
(0.30921, 0.000265692) +- (0.00310489, 7.97131e-06)
(0.324333, 0.000252559) +- (0.00298127, 6.39693e-06)
(0.354978, 0.000228848) +- (0.00347603, 5.32119e-06)
(0.384942, 0.000197438) +- (0.00361235, 4.00158e-06)
(0.398999, 0.000184615) +- (0.00345939, 3.98696e-06)
(0.40933, 0.000175357) +- (0.00336425, 4.08102e-06)
(0.416517, 0.000166286) +- (0.0034473, 3.29707e-06)
(0.436654, 0.00015003) +- (0.00333773, 3.59373e-06)
(0.451325, 0.000136927) +- (0.00334742, 2.75827e-06)
(0.461991, 0.000127882) +- (0.00310457, 2.51499e-06)
(0.472415, 0.000119369) +- (0.0031121, 2.28556e-06)
(0.481311, 0.000113074) +- (0.00309301, 2.07171e-06)
};
\addlegendentry{proposed algorithm, mean}
\addplot[dashed]
coordinates {
(0.134342, 0.123165) +- (0.00132285, 0.00248085)
(0.195808, 0.0929509) +- (0.00155259, 0.00186111)
(0.239173, 0.0749481) +- (0.00200563, 0.00151995)
(0.26841, 0.0639652) +- (0.00204706, 0.00122138)
(0.292037, 0.0555238) +- (0.00259634, 0.00105816)
(0.30921, 0.0496873) +- (0.00310489, 0.000919126)
(0.324333, 0.045166) +- (0.00298127, 0.000845781)
(0.354978, 0.0349373) +- (0.00347603, 0.000608381)
(0.384942, 0.0269187) +- (0.00361235, 0.000492528)
(0.398999, 0.0236998) +- (0.00345939, 0.000408224)
(0.40933, 0.0214757) +- (0.00336425, 0.00038275)
(0.416517, 0.0199002) +- (0.0034473, 0.000359702)
(0.436654, 0.0165383) +- (0.00333773, 0.000302427)
(0.451325, 0.0143981) +- (0.00334742, 0.000240892)
(0.461991, 0.0129049) +- (0.00310457, 0.000213193)
(0.472415, 0.0116766) +- (0.0031121, 0.000202738)
(0.481311, 0.0107234) +- (0.00309301, 0.00018496)
};
\addlegendentry{filtered-graphs algorithm, mean}
\addplot[dotted, mybars]
coordinates {
(0.134342, 0.0200562) +- (0.00132285, 0.0106696)
(0.195808, 0.0115728) +- (0.00155259, 0.00242847)
(0.239173, 0.0112596) +- (0.00200563, 0.00473137)
(0.26841, 0.0105557) +- (0.00204706, 0.00386137)
(0.292037, 0.00574618) +- (0.00259634, 0.00123907)
(0.30921, 0.00505772) +- (0.00310489, 0.00117605)
(0.324333, 0.00477432) +- (0.00298127, 0.00209182)
(0.354978, 0.0023386) +- (0.00347603, 0.000553344)
(0.384942, 0.00125079) +- (0.00361235, 0.00028527)
(0.398999, 0.00112157) +- (0.00345939, 0.000307553)
(0.40933, 0.000861458) +- (0.00336425, 0.000202511)
(0.416517, 0.000708973) +- (0.0034473, 0.000164315)
(0.436654, 0.000486876) +- (0.00333773, 9.20953e-05)
(0.451325, 0.000356712) +- (0.00334742, 5.03234e-05)
(0.461991, 0.000297423) +- (0.00310457, 4.24467e-05)
(0.472415, 0.000257559) +- (0.0031121, 3.82027e-05)
(0.481311, 0.00022813) +- (0.00309301, 3.35308e-05)
};
\addlegendentry{brute-force algorithm, mean}
\addplot[thick, solid]
coordinates {
(0.134342, 0.00204425) +- (0.00132285, 0)
(0.195808, 0.632358) +- (0.00155259, 0)
(0.239173, 0.0651908) +- (0.00200563, 0)
(0.26841, 0.600871) +- (0.00204706, 0)
(0.292037, 0.0752924) +- (0.00259634, 0)
(0.30921, 0.328941) +- (0.00310489, 0)
(0.324333, 0.0343028) +- (0.00298127, 0)
(0.354978, 0.237099) +- (0.00347603, 0)
(0.384942, 0.146819) +- (0.00361235, 0)
(0.398999, 0.347422) +- (0.00345939, 0)
(0.40933, 0.494472) +- (0.00336425, 0)
(0.416517, 0.00223229) +- (0.0034473, 0)
(0.436654, 0.689937) +- (0.00333773, 0)
(0.451325, 0.756419) +- (0.00334742, 0)
(0.461991, 0.774015) +- (0.00310457, 0)
(0.472415, 0.677969) +- (0.0031121, 0)
(0.481311, 0.663624) +- (0.00309301, 0)
};
\addlegendentry{proposed algorithm, max}
\addplot[thick, dashed]
coordinates {
(0.134342, 11.2244) +- (0.00132285, 0)
(0.195808, 0.642922) +- (0.00155259, 0)
(0.239173, 3.2055) +- (0.00200563, 0)
(0.26841, 4.49404) +- (0.00204706, 0)
(0.292037, 0.79264) +- (0.00259634, 0)
(0.30921, 0.654765) +- (0.00310489, 0)
(0.324333, 5.89729) +- (0.00298127, 0)
(0.354978, 2.08251) +- (0.00347603, 0)
(0.384942, 1.07605) +- (0.00361235, 0)
(0.398999, 2.19052) +- (0.00345939, 0)
(0.40933, 1.91204) +- (0.00336425, 0)
(0.416517, 1.92685) +- (0.0034473, 0)
(0.436654, 0.735148) +- (0.00333773, 0)
(0.451325, 0.652533) +- (0.00334742, 0)
(0.461991, 0.644015) +- (0.00310457, 0)
(0.472415, 0.645262) +- (0.0031121, 0)
(0.481311, 0.645389) +- (0.00309301, 0)
};
\addlegendentry{filtered-graphs algorithm, max}
\addplot[thick, dotted]
coordinates {
(0.134342, 201.977) +- (0.00132285, 0)
(0.195808, 25.4468) +- (0.00155259, 0)
(0.239173, 58.0454) +- (0.00200563, 0)
(0.26841, 86.239) +- (0.00204706, 0)
(0.292037, 15.3533) +- (0.00259634, 0)
(0.30921, 48.4673) +- (0.00310489, 0)
(0.324333, 107.12) +- (0.00298127, 0)
(0.354978, 32.2518) +- (0.00347603, 0)
(0.384942, 23.5027) +- (0.00361235, 0)
(0.398999, 36.771) +- (0.00345939, 0)
(0.40933, 32.5275) +- (0.00336425, 0)
(0.416517, 32.9325) +- (0.0034473, 0)
(0.436654, 14.1326) +- (0.00333773, 0)
(0.451325, 12.373) +- (0.00334742, 0)
(0.461991, 12.5113) +- (0.00310457, 0)
(0.472415, 11.8292) +- (0.0031121, 0)
(0.481311, 11.6473) +- (0.00309301, 0)
};
\addlegendentry{brute-force algorithm, max}
\end{semilogyaxis}
\end{tikzpicture}}\hfill%
  \subfloat[time taken for $\gamma = 10$, and 640 units]{%
    \label{f:time_10_640}%
    \begin{tikzpicture}
\begin{semilogyaxis}[xlabel = {network utilization}, ylabel = {time [s]}, xmin = 0.1, xmax = 0.7, height = 3.8 cm, width = 6 cm, xlabel shift = -3 pt, legend columns = 3, legend to name = regular, legend style = {/tikz/every even column/.append style = {column sep = 0.25 cm}}, ymin = 1e-5, ymax = 1e4, ylabel shift = -6 pt]
\addplot[solid]
coordinates {
(0.124004, 0.000521989) +- (0.00104124, 1.91644e-05)
(0.173418, 0.000595846) +- (0.00153556, 2.10949e-05)
(0.224284, 0.000645615) +- (0.00173251, 5.96204e-05)
(0.26456, 0.000594336) +- (0.00187045, 4.01871e-05)
(0.295828, 0.000535981) +- (0.00261182, 1.73685e-05)
(0.319363, 0.000502029) +- (0.00275961, 1.43657e-05)
(0.335841, 0.000471875) +- (0.00306959, 1.17371e-05)
(0.373713, 0.000409739) +- (0.0033594, 9.58794e-06)
(0.404553, 0.00035487) +- (0.00343917, 8.8446e-06)
(0.418391, 0.000331676) +- (0.00357912, 8.29583e-06)
(0.429609, 0.000310607) +- (0.00347659, 6.36566e-06)
(0.439332, 0.000294702) +- (0.00348815, 5.75617e-06)
(0.458518, 0.000265951) +- (0.00340261, 8.05072e-06)
(0.473956, 0.000240216) +- (0.00341648, 4.41502e-06)
(0.486677, 0.000223097) +- (0.00318962, 4.43936e-06)
(0.497379, 0.000207477) +- (0.00328742, 3.73934e-06)
(0.506702, 0.000194573) +- (0.00319273, 3.4105e-06)
};
\addlegendentry{proposed algorithm, mean}
\addplot[dashed]
coordinates {
(0.124004, 0.263132) +- (0.00104124, 0.00552606)
(0.173418, 0.208671) +- (0.00153556, 0.00498716)
(0.224284, 0.161963) +- (0.00173251, 0.00362922)
(0.26456, 0.130335) +- (0.00187045, 0.00272383)
(0.295828, 0.109141) +- (0.00261182, 0.00191492)
(0.319363, 0.0944676) +- (0.00275961, 0.00159231)
(0.335841, 0.0844463) +- (0.00306959, 0.00150753)
(0.373713, 0.0634528) +- (0.0033594, 0.00113666)
(0.404553, 0.0488007) +- (0.00343917, 0.000907774)
(0.418391, 0.043179) +- (0.00357912, 0.000805705)
(0.429609, 0.0389773) +- (0.00347659, 0.00070136)
(0.439332, 0.0356085) +- (0.00348815, 0.000650853)
(0.458518, 0.0297106) +- (0.00340261, 0.000559184)
(0.473956, 0.0257958) +- (0.00341648, 0.000463263)
(0.486677, 0.0228803) +- (0.00318962, 0.000395814)
(0.497379, 0.0208019) +- (0.00328742, 0.000367053)
(0.506702, 0.0190183) +- (0.00319273, 0.000344128)
};
\addlegendentry{filtered-graphs algorithm, mean}
\addplot[dotted, mybars]
coordinates {
(0.124004, 0.0139752) +- (0.00104124, 0.00207612)
(0.173418, 0.0119655) +- (0.00153556, 0.00179592)
(0.224284, 0.0114788) +- (0.00173251, 0.00222961)
(0.26456, 0.00876705) +- (0.00187045, 0.00175083)
(0.295828, 0.0060572) +- (0.00261182, 0.000941095)
(0.319363, 0.00551785) +- (0.00275961, 0.00174118)
(0.335841, 0.0037102) +- (0.00306959, 0.000493199)
(0.373713, 0.00217512) +- (0.0033594, 0.000331627)
(0.404553, 0.00125323) +- (0.00343917, 0.000192963)
(0.418391, 0.00104835) +- (0.00357912, 0.000202753)
(0.429609, 0.000839282) +- (0.00347659, 0.000149968)
(0.439332, 0.000683853) +- (0.00348815, 9.2241e-05)
(0.458518, 0.000559514) +- (0.00340261, 0.000118152)
(0.473956, 0.000446941) +- (0.00341648, 8.842e-05)
(0.486677, 0.000388484) +- (0.00318962, 8.40198e-05)
(0.497379, 0.000341526) +- (0.00328742, 8.10911e-05)
(0.506702, 0.000293538) +- (0.00319273, 6.17036e-05)
};
\addlegendentry{brute-force algorithm, mean}
\addplot[thick, solid]
coordinates {
(0.124004, 0.5479) +- (0.00104124, 0)
(0.173418, 0.579815) +- (0.00153556, 0)
(0.224284, 1.96522) +- (0.00173251, 0)
(0.26456, 2.8179) +- (0.00187045, 0)
(0.295828, 1.02719) +- (0.00261182, 0)
(0.319363, 0.848007) +- (0.00275961, 0)
(0.335841, 0.0321254) +- (0.00306959, 0)
(0.373713, 0.313629) +- (0.0033594, 0)
(0.404553, 0.857246) +- (0.00343917, 0)
(0.418391, 1.74161) +- (0.00357912, 0)
(0.429609, 0.455699) +- (0.00347659, 0)
(0.439332, 0.307627) +- (0.00348815, 0)
(0.458518, 4.85399) +- (0.00340261, 0)
(0.473956, 0.922413) +- (0.00341648, 0)
(0.486677, 2.3888) +- (0.00318962, 0)
(0.497379, 0.893268) +- (0.00328742, 0)
(0.506702, 1.90049) +- (0.00319273, 0)
};
\addlegendentry{proposed algorithm, max}
\addplot[thick, dashed]
coordinates {
(0.124004, 1.38777) +- (0.00104124, 0)
(0.173418, 1.38541) +- (0.00153556, 0)
(0.224284, 1.38079) +- (0.00173251, 0)
(0.26456, 1.5499) +- (0.00187045, 0)
(0.295828, 1.39041) +- (0.00261182, 0)
(0.319363, 9.91664) +- (0.00275961, 0)
(0.335841, 1.40038) +- (0.00306959, 0)
(0.373713, 1.38338) +- (0.0033594, 0)
(0.404553, 1.38265) +- (0.00343917, 0)
(0.418391, 1.76975) +- (0.00357912, 0)
(0.429609, 1.38515) +- (0.00347659, 0)
(0.439332, 1.74108) +- (0.00348815, 0)
(0.458518, 1.38425) +- (0.00340261, 0)
(0.473956, 4.09701) +- (0.00341648, 0)
(0.486677, 3.80949) +- (0.00318962, 0)
(0.497379, 3.82272) +- (0.00328742, 0)
(0.506702, 4.1145) +- (0.00319273, 0)
};
\addlegendentry{filtered-graphs algorithm, max}
\addplot[thick, dotted]
coordinates {
(0.124004, 15.6002) +- (0.00104124, 0)
(0.173418, 23.6941) +- (0.00153556, 0)
(0.224284, 37.6357) +- (0.00173251, 0)
(0.26456, 53.4557) +- (0.00187045, 0)
(0.295828, 17.3338) +- (0.00261182, 0)
(0.319363, 182.754) +- (0.00275961, 0)
(0.335841, 7.96392) +- (0.00306959, 0)
(0.373713, 15.0148) +- (0.0033594, 0)
(0.404553, 16.7438) +- (0.00343917, 0)
(0.418391, 31.7569) +- (0.00357912, 0)
(0.429609, 12.5498) +- (0.00347659, 0)
(0.439332, 28.7368) +- (0.00348815, 0)
(0.458518, 73.145) +- (0.00340261, 0)
(0.473956, 59.5987) +- (0.00341648, 0)
(0.486677, 57.8726) +- (0.00318962, 0)
(0.497379, 56.9278) +- (0.00328742, 0)
(0.506702, 59.3093) +- (0.00319273, 0)
};
\addlegendentry{brute-force algorithm, max}
\end{semilogyaxis}
\end{tikzpicture}}\\[3pt]
  \subfloat[memory used for $\gamma = 1$, and 160 units]{%
    \label{f:memory_1_160}%
    \begin{tikzpicture}
\begin{semilogyaxis}[xlabel = {network utilization}, ylabel = {memory words}, xmin = 0.1, xmax = 0.7, height = 3.8 cm, width = 6 cm, xlabel shift = -3 pt, legend columns = 3, legend to name = regular, legend style = {/tikz/every even column/.append style = {column sep = 0.25 cm}}, ymin = 1e1, ymax = 1e11, ylabel shift = -4 pt]
\addplot[solid]
coordinates {
(0.131461, 3449.624) +- (0.00071791, 66.4887)
(0.191483, 4182.077) +- (0.00112687, 84.2848)
(0.259328, 4174.401) +- (0.00218623, 101.8353)
(0.323594, 3663.752) +- (0.00252102, 85.2198)
(0.363878, 3285.702) +- (0.00283414, 63.5927)
(0.390379, 3025.9) +- (0.00337979, 53.475)
(0.409166, 2847.873) +- (0.00376252, 45.579)
(0.4551, 2429.598) +- (0.0042842, 39.523)
(0.491548, 2120.155) +- (0.00443894, 33.50701)
(0.508246, 1988.262) +- (0.00465782, 30.43196)
(0.522032, 1887.658) +- (0.00459475, 29.09627)
(0.533248, 1805.465) +- (0.00447518, 28.49624)
(0.556131, 1649.145) +- (0.00432393, 24.95293)
(0.574202, 1535.575) +- (0.00414095, 22.6994)
(0.589254, 1444.73) +- (0.00398702, 20.73229)
(0.60164, 1373.528) +- (0.0040399, 19.70865)
(0.612831, 1309.533) +- (0.00389073, 18.22082)
};
\addlegendentry{proposed algorithm, mean}
\addplot[dashed]
coordinates {
(0.131461, 159.9514) +- (0.00071791, 0.677625)
(0.191483, 156.6995) +- (0.00112687, 0.835696)
(0.259328, 145.1846) +- (0.00218623, 1.288118)
(0.323594, 129.2871) +- (0.00252102, 1.204681)
(0.363878, 117.8759) +- (0.00283414, 0.770869)
(0.390379, 109.6232) +- (0.00337979, 0.605239)
(0.409166, 103.5196) +- (0.00376252, 0.538745)
(0.4551, 87.8201) +- (0.0042842, 0.385745)
(0.491548, 74.9972) +- (0.00443894, 0.359033)
(0.508246, 69.3513) +- (0.00465782, 0.375193)
(0.522032, 65.1261) +- (0.00459475, 0.361915)
(0.533248, 61.7014) +- (0.00447518, 0.379079)
(0.556131, 55.4779) +- (0.00432393, 0.349162)
(0.574202, 51.195) +- (0.00414095, 0.334069)
(0.589254, 47.9192) +- (0.00398702, 0.321729)
(0.60164, 45.328) +- (0.0040399, 0.31505)
(0.612831, 43.2918) +- (0.00389073, 0.2966645)
};
\addlegendentry{filtered-graphs algorithm, mean}
\addplot[dotted, mybars]
coordinates {
(0.131461, 119584.49) +- (0.00071791, 15219.614)
(0.191483, 103166.62) +- (0.00112687, 12703.098)
(0.259328, 78451.25) +- (0.00218623, 11542.25)
(0.323594, 45329.24) +- (0.00252102, 6462.633)
(0.363878, 31976.19) +- (0.00283414, 5284.532)
(0.390379, 23757.95) +- (0.00337979, 3644.288)
(0.409166, 19024.748) +- (0.00376252, 3001.2977)
(0.4551, 9599.949) +- (0.0042842, 3625.639)
(0.491548, 7946.427) +- (0.00443894, 2360.667)
(0.508246, 5528.688) +- (0.00465782, 1054.4271)
(0.522032, 5489.881) +- (0.00459475, 1509.1035)
(0.533248, 4737.223) +- (0.00447518, 1734.325)
(0.556131, 3214.208) +- (0.00432393, 932.0447)
(0.574202, 2486.3664) +- (0.00414095, 564.184)
(0.589254, 2170.6245) +- (0.00398702, 707.5863)
(0.60164, 1618.6185) +- (0.0040399, 259.85929)
(0.612831, 1426.8679) +- (0.00389073, 638.1294)
};
\addlegendentry{brute-force algorithm, mean}
\addplot[thick, solid]
coordinates {
(0.131461, 16810.0) +- (0.00071791, 0)
(0.191483, 18385.0) +- (0.00112687, 0)
(0.259328, 18675.0) +- (0.00218623, 0)
(0.323594, 16595.0) +- (0.00252102, 0)
(0.363878, 15895.0) +- (0.00283414, 0)
(0.390379, 16225.0) +- (0.00337979, 0)
(0.409166, 17725.0) +- (0.00376252, 0)
(0.4551, 15990.0) +- (0.0042842, 0)
(0.491548, 17700.0) +- (0.00443894, 0)
(0.508246, 15880.0) +- (0.00465782, 0)
(0.522032, 14810.0) +- (0.00459475, 0)
(0.533248, 17195.0) +- (0.00447518, 0)
(0.556131, 15185.0) +- (0.00432393, 0)
(0.574202, 16670.0) +- (0.00414095, 0)
(0.589254, 15000.0) +- (0.00398702, 0)
(0.60164, 16650.0) +- (0.0040399, 0)
(0.612831, 17355.0) +- (0.00389073, 0)
};
\addlegendentry{proposed algorithm, max}
\addplot[thick, dashed]
coordinates {
(0.131461, 296.0) +- (0.00071791, 0)
(0.191483, 290.0) +- (0.00112687, 0)
(0.259328, 290.0) +- (0.00218623, 0)
(0.323594, 290.0) +- (0.00252102, 0)
(0.363878, 290.0) +- (0.00283414, 0)
(0.390379, 290.0) +- (0.00337979, 0)
(0.409166, 290.0) +- (0.00376252, 0)
(0.4551, 290.0) +- (0.0042842, 0)
(0.491548, 290.0) +- (0.00443894, 0)
(0.508246, 290.0) +- (0.00465782, 0)
(0.522032, 290.0) +- (0.00459475, 0)
(0.533248, 290.0) +- (0.00447518, 0)
(0.556131, 290.0) +- (0.00432393, 0)
(0.574202, 290.0) +- (0.00414095, 0)
(0.589254, 290.0) +- (0.00398702, 0)
(0.60164, 290.0) +- (0.0040399, 0)
(0.612831, 290.0) +- (0.00389073, 0)
};
\addlegendentry{filtered-graphs algorithm, max}
\addplot[thick, dotted]
coordinates {
(0.131461, 59320270.0) +- (0.00071791, 0)
(0.191483, 56881100.0) +- (0.00112687, 0)
(0.259328, 390523700.0) +- (0.00218623, 0)
(0.323594, 115784800.0) +- (0.00252102, 0)
(0.363878, 183401180.0) +- (0.00283414, 0)
(0.390379, 508204900.0) +- (0.00337979, 0)
(0.409166, 258315680.0) +- (0.00376252, 0)
(0.4551, 412860400.0) +- (0.0042842, 0)
(0.491548, 1124563800.0) +- (0.00443894, 0)
(0.508246, 270117770.0) +- (0.00465782, 0)
(0.522032, 1099948800.0) +- (0.00459475, 0)
(0.533248, 1878069700.0) +- (0.00447518, 0)
(0.556131, 1134802700.0) +- (0.00432393, 0)
(0.574202, 757700000.0) +- (0.00414095, 0)
(0.589254, 1390835600.0) +- (0.00398702, 0)
(0.60164, 131388240.0) +- (0.0040399, 0)
(0.612831, 1278506700.0) +- (0.00389073, 0)
};
\addlegendentry{brute-force algorithm, max}
\end{semilogyaxis}
\end{tikzpicture}}\hfill%
  \subfloat[memory used for $\gamma = 1$, and 320 units]{%
    \label{f:memory_1_320}%
    \begin{tikzpicture}
\begin{semilogyaxis}[xlabel = {network utilization}, ylabel = {memory words}, xmin = 0.1, xmax = 0.7, height = 3.8 cm, width = 6 cm, xlabel shift = -3 pt, legend columns = 3, legend to name = regular, legend style = {/tikz/every even column/.append style = {column sep = 0.25 cm}}, ymin = 1e1, ymax = 1e11, ylabel shift = -4 pt]
\addplot[solid]
coordinates {
(0.127142, 5689.0) +- (0.000588861, 119.9699)
(0.19061, 7425.77) +- (0.000927054, 151.5121)
(0.253873, 7752.97) +- (0.00161753, 179.0396)
(0.320806, 6936.12) +- (0.00220855, 170.3891)
(0.370566, 6055.17) +- (0.00265177, 125.5844)
(0.401078, 5521.45) +- (0.00336707, 101.8692)
(0.421552, 5166.12) +- (0.00382383, 91.0978)
(0.469919, 4380.749) +- (0.00453692, 74.746)
(0.50895, 3800.299) +- (0.00480454, 65.033)
(0.526206, 3564.073) +- (0.00483655, 60.26)
(0.53985, 3381.326) +- (0.00486905, 58.733)
(0.551682, 3233.049) +- (0.00476024, 53.6778)
(0.574532, 2958.551) +- (0.00456749, 48.80128)
(0.59286, 2754.499) +- (0.00435369, 44.38652)
(0.608082, 2590.1) +- (0.00420392, 40.09927)
(0.621302, 2456.172) +- (0.00412613, 36.93644)
(0.632771, 2341.438) +- (0.00406718, 34.7195)
};
\addlegendentry{proposed algorithm, mean}
\addplot[dashed]
coordinates {
(0.127142, 160.9955) +- (0.000588861, 0.647136)
(0.19061, 160.259) +- (0.000927054, 0.728765)
(0.253873, 151.3183) +- (0.00161753, 1.215803)
(0.320806, 135.2958) +- (0.00220855, 1.424537)
(0.370566, 121.4538) +- (0.00265177, 1.033439)
(0.401078, 112.069) +- (0.00336707, 0.686199)
(0.421552, 105.5295) +- (0.00382383, 0.56264)
(0.469919, 88.8833) +- (0.00453692, 0.400737)
(0.50895, 75.4478) +- (0.00480454, 0.389177)
(0.526206, 69.7751) +- (0.00483655, 0.399497)
(0.53985, 65.3039) +- (0.00486905, 0.411464)
(0.551682, 61.9079) +- (0.00476024, 0.40615)
(0.574532, 55.6028) +- (0.00456749, 0.39517)
(0.59286, 51.3553) +- (0.00435369, 0.370071)
(0.608082, 48.0434) +- (0.00420392, 0.344498)
(0.621302, 45.5108) +- (0.00412613, 0.327366)
(0.632771, 43.422) +- (0.00406718, 0.311434)
};
\addlegendentry{filtered-graphs algorithm, mean}
\addplot[dotted, mybars]
coordinates {
(0.127142, 133921.6) +- (0.000588861, 16593.141)
(0.19061, 128824.92) +- (0.000927054, 15531.749)
(0.253873, 98300.32) +- (0.00161753, 15114.963)
(0.320806, 59878.26) +- (0.00220855, 8843.236)
(0.370566, 38573.27) +- (0.00265177, 5999.878)
(0.401078, 27753.178) +- (0.00336707, 3729.494)
(0.421552, 23964.73) +- (0.00382383, 5835.646)
(0.469919, 12796.277) +- (0.00453692, 2062.9151)
(0.50895, 7407.088) +- (0.00480454, 945.9389)
(0.526206, 6060.246) +- (0.00483655, 825.211)
(0.53985, 5280.418) +- (0.00486905, 890.3592)
(0.551682, 4284.372) +- (0.00476024, 561.8772)
(0.574532, 3229.088) +- (0.00456749, 431.4489)
(0.59286, 2597.8732) +- (0.00435369, 408.8309)
(0.608082, 2119.466) +- (0.00420392, 231.17876)
(0.621302, 1929.3857) +- (0.00412613, 242.00654)
(0.632771, 1719.8167) +- (0.00406718, 221.95485)
};
\addlegendentry{brute-force algorithm, mean}
\addplot[thick, solid]
coordinates {
(0.127142, 28065.0) +- (0.000588861, 0)
(0.19061, 34515.0) +- (0.000927054, 0)
(0.253873, 36275.0) +- (0.00161753, 0)
(0.320806, 33995.0) +- (0.00220855, 0)
(0.370566, 30440.0) +- (0.00265177, 0)
(0.401078, 29920.0) +- (0.00336707, 0)
(0.421552, 29845.0) +- (0.00382383, 0)
(0.469919, 32125.0) +- (0.00453692, 0)
(0.50895, 30820.0) +- (0.00480454, 0)
(0.526206, 30955.0) +- (0.00483655, 0)
(0.53985, 31320.0) +- (0.00486905, 0)
(0.551682, 27990.0) +- (0.00476024, 0)
(0.574532, 30725.0) +- (0.00456749, 0)
(0.59286, 27205.0) +- (0.00435369, 0)
(0.608082, 28785.0) +- (0.00420392, 0)
(0.621302, 29010.0) +- (0.00412613, 0)
(0.632771, 29835.0) +- (0.00406718, 0)
};
\addlegendentry{proposed algorithm, max}
\addplot[thick, dashed]
coordinates {
(0.127142, 296.0) +- (0.000588861, 0)
(0.19061, 296.0) +- (0.000927054, 0)
(0.253873, 296.0) +- (0.00161753, 0)
(0.320806, 296.0) +- (0.00220855, 0)
(0.370566, 296.0) +- (0.00265177, 0)
(0.401078, 296.0) +- (0.00336707, 0)
(0.421552, 296.0) +- (0.00382383, 0)
(0.469919, 290.0) +- (0.00453692, 0)
(0.50895, 290.0) +- (0.00480454, 0)
(0.526206, 290.0) +- (0.00483655, 0)
(0.53985, 290.0) +- (0.00486905, 0)
(0.551682, 290.0) +- (0.00476024, 0)
(0.574532, 290.0) +- (0.00456749, 0)
(0.59286, 290.0) +- (0.00435369, 0)
(0.608082, 290.0) +- (0.00420392, 0)
(0.621302, 290.0) +- (0.00412613, 0)
(0.632771, 290.0) +- (0.00406718, 0)
};
\addlegendentry{filtered-graphs algorithm, max}
\addplot[thick, dotted]
coordinates {
(0.127142, 62730680.0) +- (0.000588861, 0)
(0.19061, 128313230.0) +- (0.000927054, 0)
(0.253873, 2330414300.0) +- (0.00161753, 0)
(0.320806, 183372790.0) +- (0.00220855, 0)
(0.370566, 270803240.0) +- (0.00265177, 0)
(0.401078, 319510180.0) +- (0.00336707, 0)
(0.421552, 2143528000.0) +- (0.00382383, 0)
(0.469919, 1226766000.0) +- (0.00453692, 0)
(0.50895, 220067660.0) +- (0.00480454, 0)
(0.526206, 195103220.0) +- (0.00483655, 0)
(0.53985, 873901200.0) +- (0.00486905, 0)
(0.551682, 182192590.0) +- (0.00476024, 0)
(0.574532, 398434780.0) +- (0.00456749, 0)
(0.59286, 514108800.0) +- (0.00435369, 0)
(0.608082, 157883530.0) +- (0.00420392, 0)
(0.621302, 381576340.0) +- (0.00412613, 0)
(0.632771, 390358630.0) +- (0.00406718, 0)
};
\addlegendentry{brute-force algorithm, max}
\end{semilogyaxis}
\end{tikzpicture}}\hfill%
  \subfloat[memory used for $\gamma = 1$, and 640 units]{%
    \label{f:memory_1_640}%
    \begin{tikzpicture}
\begin{semilogyaxis}[xlabel = {network utilization}, ylabel = {memory words}, xmin = 0.1, xmax = 0.7, height = 3.8 cm, width = 6 cm, xlabel shift = -3 pt, legend columns = 3, legend to name = regular, legend style = {/tikz/every even column/.append style = {column sep = 0.25 cm}}, ymin = 1e1, ymax = 1e11, ylabel shift = -4 pt]
\addplot[solid]
coordinates {
(0.125972, 9708.45) +- (0.000588153, 204.8861)
(0.189211, 13177.78) +- (0.000881459, 278.4199)
(0.254461, 14351.28) +- (0.00136103, 324.2676)
(0.319807, 13154.05) +- (0.00209886, 326.1926)
(0.3748, 11381.6) +- (0.00262717, 246.0638)
(0.40838, 10262.65) +- (0.00343446, 199.171)
(0.430671, 9569.82) +- (0.00395443, 182.3288)
(0.481924, 8039.55) +- (0.00470082, 152.1914)
(0.52315, 6937.4) +- (0.00497199, 129.5366)
(0.540775, 6503.37) +- (0.00512155, 121.7777)
(0.555105, 6172.55) +- (0.00507492, 114.8381)
(0.566632, 5913.97) +- (0.0049505, 105.8981)
(0.589688, 5421.68) +- (0.0047501, 95.8625)
(0.608263, 5037.72) +- (0.0045344, 84.7517)
(0.623658, 4737.251) +- (0.00438106, 78.2967)
(0.636988, 4494.228) +- (0.00429597, 71.9583)
(0.648684, 4278.127) +- (0.00424394, 67.3482)
};
\addlegendentry{proposed algorithm, mean}
\addplot[dashed]
coordinates {
(0.125972, 161.8069) +- (0.000588153, 0.634153)
(0.189211, 162.0816) +- (0.000881459, 0.676277)
(0.254461, 155.2178) +- (0.00136103, 1.143309)
(0.319807, 140.3408) +- (0.00209886, 1.483765)
(0.3748, 124.9682) +- (0.00262717, 1.162134)
(0.40838, 114.6158) +- (0.00343446, 0.757182)
(0.430671, 107.3932) +- (0.00395443, 0.579299)
(0.481924, 89.7622) +- (0.00470082, 0.445342)
(0.52315, 75.9302) +- (0.00497199, 0.437597)
(0.540775, 70.1067) +- (0.00512155, 0.437949)
(0.555105, 65.7088) +- (0.00507492, 0.449764)
(0.566632, 62.2477) +- (0.0049505, 0.442892)
(0.589688, 55.9449) +- (0.0047501, 0.425813)
(0.608263, 51.5342) +- (0.0045344, 0.394915)
(0.623658, 48.274) +- (0.00438106, 0.369374)
(0.636988, 45.7192) +- (0.00429597, 0.347187)
(0.648684, 43.6522) +- (0.00424394, 0.331725)
};
\addlegendentry{filtered-graphs algorithm, mean}
\addplot[dotted, mybars]
coordinates {
(0.125972, 153991.32) +- (0.000588153, 18703.02)
(0.189211, 164892.66) +- (0.000881459, 19764.525)
(0.254461, 126914.18) +- (0.00136103, 16004.195)
(0.319807, 85475.66) +- (0.00209886, 15907.499)
(0.3748, 52721.71) +- (0.00262717, 9199.788)
(0.40838, 36655.58) +- (0.00343446, 5350.607)
(0.430671, 28752.688) +- (0.00395443, 4305.364)
(0.481924, 16307.508) +- (0.00470082, 2452.7867)
(0.52315, 10374.78) +- (0.00497199, 1815.8958)
(0.540775, 8024.011) +- (0.00512155, 1287.933)
(0.555105, 6476.051) +- (0.00507492, 923.6344)
(0.566632, 5734.455) +- (0.0049505, 936.4373)
(0.589688, 4413.206) +- (0.0047501, 736.7457)
(0.608263, 3619.3285) +- (0.0045344, 640.284)
(0.623658, 3032.1476) +- (0.00438106, 469.9543)
(0.636988, 2614.3502) +- (0.00429597, 346.29706)
(0.648684, 2393.5244) +- (0.00424394, 347.24276)
};
\addlegendentry{brute-force algorithm, mean}
\addplot[thick, solid]
coordinates {
(0.125972, 45420.0) +- (0.000588153, 0)
(0.189211, 61910.0) +- (0.000881459, 0)
(0.254461, 67070.0) +- (0.00136103, 0)
(0.319807, 59350.0) +- (0.00209886, 0)
(0.3748, 58205.0) +- (0.00262717, 0)
(0.40838, 62480.0) +- (0.00343446, 0)
(0.430671, 58840.0) +- (0.00395443, 0)
(0.481924, 59735.0) +- (0.00470082, 0)
(0.52315, 59385.0) +- (0.00497199, 0)
(0.540775, 59800.0) +- (0.00512155, 0)
(0.555105, 56375.0) +- (0.00507492, 0)
(0.566632, 56540.0) +- (0.0049505, 0)
(0.589688, 64155.0) +- (0.0047501, 0)
(0.608263, 58445.0) +- (0.0045344, 0)
(0.623658, 56615.0) +- (0.00438106, 0)
(0.636988, 58640.0) +- (0.00429597, 0)
(0.648684, 55980.0) +- (0.00424394, 0)
};
\addlegendentry{proposed algorithm, max}
\addplot[thick, dashed]
coordinates {
(0.125972, 296.0) +- (0.000588153, 0)
(0.189211, 296.0) +- (0.000881459, 0)
(0.254461, 296.0) +- (0.00136103, 0)
(0.319807, 296.0) +- (0.00209886, 0)
(0.3748, 296.0) +- (0.00262717, 0)
(0.40838, 296.0) +- (0.00343446, 0)
(0.430671, 296.0) +- (0.00395443, 0)
(0.481924, 296.0) +- (0.00470082, 0)
(0.52315, 296.0) +- (0.00497199, 0)
(0.540775, 296.0) +- (0.00512155, 0)
(0.555105, 296.0) +- (0.00507492, 0)
(0.566632, 296.0) +- (0.0049505, 0)
(0.589688, 296.0) +- (0.0047501, 0)
(0.608263, 296.0) +- (0.0045344, 0)
(0.623658, 296.0) +- (0.00438106, 0)
(0.636988, 296.0) +- (0.00429597, 0)
(0.648684, 296.0) +- (0.00424394, 0)
};
\addlegendentry{filtered-graphs algorithm, max}
\addplot[thick, dotted]
coordinates {
(0.125972, 72119750.0) +- (0.000588153, 0)
(0.189211, 138271860.0) +- (0.000881459, 0)
(0.254461, 542357300.0) +- (0.00136103, 0)
(0.319807, 242169140.0) +- (0.00209886, 0)
(0.3748, 412659180.0) +- (0.00262717, 0)
(0.40838, 265608400.0) +- (0.00343446, 0)
(0.430671, 169376950.0) +- (0.00395443, 0)
(0.481924, 1195675400.0) +- (0.00470082, 0)
(0.52315, 1286991000.0) +- (0.00497199, 0)
(0.540775, 797338500.0) +- (0.00512155, 0)
(0.555105, 536873800.0) +- (0.00507492, 0)
(0.566632, 725224900.0) +- (0.0049505, 0)
(0.589688, 566064800.0) +- (0.0047501, 0)
(0.608263, 584765100.0) +- (0.0045344, 0)
(0.623658, 462763720.0) +- (0.00438106, 0)
(0.636988, 388379750.0) +- (0.00429597, 0)
(0.648684, 699057700.0) +- (0.00424394, 0)
};
\addlegendentry{brute-force algorithm, max}
\end{semilogyaxis}
\end{tikzpicture}}\\[3pt]
  \subfloat[memory used for $\gamma = 10$, and 160 units]{%
    \label{f:memory_10_160}%
    \begin{tikzpicture}
\begin{semilogyaxis}[xlabel = {network utilization}, ylabel = {memory words}, xmin = 0.1, xmax = 0.7, height = 3.8 cm, width = 6 cm, xlabel shift = -3 pt, legend columns = 3, legend to name = regular, legend style = {/tikz/every even column/.append style = {column sep = 0.25 cm}}, ymin = 1e1, ymax = 1e11, ylabel shift = -4 pt]
\addplot[solid]
coordinates {
(0.123634, 662.6) +- (0.0017087, 14.00027)
(0.194931, 696.435) +- (0.00286287, 15.1787)
(0.236609, 674.562) +- (0.00297472, 14.21735)
(0.259264, 655.505) +- (0.00314873, 13.17158)
(0.276294, 628.353) +- (0.00342659, 12.82591)
(0.291206, 609.35) +- (0.0031663, 11.24097)
(0.301845, 584.367) +- (0.00336988, 10.61315)
(0.333706, 523.258) +- (0.00347518, 8.54812)
(0.362026, 461.7473) +- (0.00357163, 7.62218)
(0.374476, 435.7548) +- (0.00358651, 6.18142)
(0.383973, 415.7652) +- (0.00353857, 6.06868)
(0.391935, 396.1599) +- (0.00352054, 5.7528)
(0.411353, 355.5076) +- (0.00341638, 4.930294)
(0.424501, 330.5223) +- (0.00327267, 4.327601)
(0.43423, 309.0627) +- (0.00307809, 3.957705)
(0.443782, 291.7524) +- (0.00323497, 3.57827)
(0.453735, 275.3724) +- (0.00314611, 3.263446)
};
\addlegendentry{proposed algorithm, mean}
\addplot[dashed]
coordinates {
(0.123634, 128.476) +- (0.0017087, 0.859615)
(0.194931, 118.4177) +- (0.00286287, 0.744847)
(0.236609, 111.1949) +- (0.00297472, 0.680818)
(0.259264, 105.502) +- (0.00314873, 0.630372)
(0.276294, 100.0941) +- (0.00342659, 0.552802)
(0.291206, 96.1489) +- (0.0031663, 0.54687)
(0.301845, 91.8239) +- (0.00336988, 0.544851)
(0.333706, 80.6621) +- (0.00347518, 0.446773)
(0.362026, 70.1779) +- (0.00357163, 0.376163)
(0.374476, 65.7541) +- (0.00358651, 0.344661)
(0.383973, 62.3393) +- (0.00353857, 0.369402)
(0.391935, 59.3713) +- (0.00352054, 0.370142)
(0.411353, 53.2147) +- (0.00341638, 0.312681)
(0.424501, 49.4247) +- (0.00327267, 0.2917609)
(0.43423, 46.383) +- (0.00307809, 0.2860481)
(0.443782, 44.0421) +- (0.00323497, 0.2700385)
(0.453735, 41.8969) +- (0.00314611, 0.2641579)
};
\addlegendentry{filtered-graphs algorithm, mean}
\addplot[dotted, mybars]
coordinates {
(0.123634, 96701.06) +- (0.0017087, 14705.749)
(0.194931, 79748.71) +- (0.00286287, 20037.496)
(0.236609, 47623.36) +- (0.00297472, 8133.75)
(0.259264, 40265.54) +- (0.00314873, 11531.105)
(0.276294, 35326.31) +- (0.00342659, 8938.145)
(0.291206, 23853.686) +- (0.0031663, 4114.906)
(0.301845, 20872.391) +- (0.00336988, 5290.782)
(0.333706, 11848.757) +- (0.00347518, 2885.271)
(0.362026, 5787.831) +- (0.00357163, 989.0898)
(0.374476, 6070.348) +- (0.00358651, 3406.5364)
(0.383973, 4742.663) +- (0.00353857, 2836.9229)
(0.391935, 4126.571) +- (0.00352054, 2523.3762)
(0.411353, 2981.605) +- (0.00341638, 1886.113)
(0.424501, 2454.5468) +- (0.00327267, 1558.6836)
(0.43423, 1908.8953) +- (0.00307809, 1263.7112)
(0.443782, 1613.8997) +- (0.00323497, 1089.9518)
(0.453735, 1409.2491) +- (0.00314611, 944.0236)
};
\addlegendentry{brute-force algorithm, mean}
\addplot[thick, solid]
coordinates {
(0.123634, 3420.0) +- (0.0017087, 0)
(0.194931, 3965.0) +- (0.00286287, 0)
(0.236609, 4305.0) +- (0.00297472, 0)
(0.259264, 4235.0) +- (0.00314873, 0)
(0.276294, 4620.0) +- (0.00342659, 0)
(0.291206, 4245.0) +- (0.0031663, 0)
(0.301845, 4420.0) +- (0.00336988, 0)
(0.333706, 3960.0) +- (0.00347518, 0)
(0.362026, 5045.0) +- (0.00357163, 0)
(0.374476, 4110.0) +- (0.00358651, 0)
(0.383973, 4875.0) +- (0.00353857, 0)
(0.391935, 4420.0) +- (0.00352054, 0)
(0.411353, 4675.0) +- (0.00341638, 0)
(0.424501, 4230.0) +- (0.00327267, 0)
(0.43423, 3755.0) +- (0.00307809, 0)
(0.443782, 4510.0) +- (0.00323497, 0)
(0.453735, 4255.0) +- (0.00314611, 0)
};
\addlegendentry{proposed algorithm, max}
\addplot[thick, dashed]
coordinates {
(0.123634, 284.0) +- (0.0017087, 0)
(0.194931, 284.0) +- (0.00286287, 0)
(0.236609, 284.0) +- (0.00297472, 0)
(0.259264, 284.0) +- (0.00314873, 0)
(0.276294, 284.0) +- (0.00342659, 0)
(0.291206, 284.0) +- (0.0031663, 0)
(0.301845, 284.0) +- (0.00336988, 0)
(0.333706, 284.0) +- (0.00347518, 0)
(0.362026, 284.0) +- (0.00357163, 0)
(0.374476, 284.0) +- (0.00358651, 0)
(0.383973, 284.0) +- (0.00353857, 0)
(0.391935, 284.0) +- (0.00352054, 0)
(0.411353, 284.0) +- (0.00341638, 0)
(0.424501, 284.0) +- (0.00327267, 0)
(0.43423, 284.0) +- (0.00307809, 0)
(0.443782, 284.0) +- (0.00323497, 0)
(0.453735, 284.0) +- (0.00314611, 0)
};
\addlegendentry{filtered-graphs algorithm, max}
\addplot[thick, dotted]
coordinates {
(0.123634, 72646450.0) +- (0.0017087, 0)
(0.194931, 106799710.0) +- (0.00286287, 0)
(0.236609, 40568295.0) +- (0.00297472, 0)
(0.259264, 150693970.0) +- (0.00314873, 0)
(0.276294, 85099290.0) +- (0.00342659, 0)
(0.291206, 48301350.0) +- (0.0031663, 0)
(0.301845, 88718690.0) +- (0.00336988, 0)
(0.333706, 61654720.0) +- (0.00347518, 0)
(0.362026, 20829388.0) +- (0.00357163, 0)
(0.374476, 363177480.0) +- (0.00358651, 0)
(0.383973, 360505000.0) +- (0.00353857, 0)
(0.391935, 360493080.0) +- (0.00352054, 0)
(0.411353, 360493080.0) +- (0.00341638, 0)
(0.424501, 360493080.0) +- (0.00327267, 0)
(0.43423, 360493080.0) +- (0.00307809, 0)
(0.443782, 360493080.0) +- (0.00323497, 0)
(0.453735, 360493080.0) +- (0.00314611, 0)
};
\addlegendentry{brute-force algorithm, max}
\end{semilogyaxis}
\end{tikzpicture}}\hfill%
  \subfloat[memory used for $\gamma = 10$, and 320 units]{%
    \label{f:memory_10_320}%
    \begin{tikzpicture}
\begin{semilogyaxis}[xlabel = {network utilization}, ylabel = {memory words}, xmin = 0.1, xmax = 0.7, height = 3.8 cm, width = 6 cm, xlabel shift = -3 pt, legend columns = 3, legend to name = regular, legend style = {/tikz/every even column/.append style = {column sep = 0.25 cm}}, ymin = 1e1, ymax = 1e11, ylabel shift = -4 pt]
\addplot[solid]
coordinates {
(0.134342, 1196.048) +- (0.00132285, 25.45426)
(0.195808, 1280.718) +- (0.00155259, 29.59399)
(0.239173, 1224.658) +- (0.00200563, 27.87649)
(0.26841, 1163.793) +- (0.00204706, 24.90723)
(0.292037, 1099.257) +- (0.00259634, 22.47038)
(0.30921, 1041.31) +- (0.00310489, 20.01046)
(0.324333, 999.69) +- (0.00298127, 19.69973)
(0.354978, 902.03) +- (0.00347603, 14.8735)
(0.384942, 782.548) +- (0.00361235, 12.08834)
(0.398999, 732.62) +- (0.00345939, 10.88367)
(0.40933, 693.475) +- (0.00336425, 9.99702)
(0.416517, 662.555) +- (0.0034473, 9.47881)
(0.436654, 595.102) +- (0.00333773, 8.23389)
(0.451325, 548.088) +- (0.00334742, 7.46633)
(0.461991, 514.265) +- (0.00310457, 6.62624)
(0.472415, 483.2148) +- (0.0031121, 6.45227)
(0.481311, 460.3748) +- (0.00309301, 5.76636)
};
\addlegendentry{proposed algorithm, mean}
\addplot[dashed]
coordinates {
(0.134342, 132.8483) +- (0.00132285, 0.670183)
(0.195808, 122.8862) +- (0.00155259, 0.621305)
(0.239173, 113.6684) +- (0.00200563, 0.553577)
(0.26841, 106.98) +- (0.00204706, 0.512569)
(0.292037, 100.7254) +- (0.00259634, 0.518784)
(0.30921, 95.5223) +- (0.00310489, 0.459284)
(0.324333, 91.5551) +- (0.00298127, 0.449722)
(0.354978, 80.959) +- (0.00347603, 0.350015)
(0.384942, 70.2144) +- (0.00361235, 0.3956)
(0.398999, 65.5031) +- (0.00345939, 0.367063)
(0.40933, 61.9092) +- (0.00336425, 0.343132)
(0.416517, 59.1179) +- (0.0034473, 0.328861)
(0.436654, 53.4105) +- (0.00333773, 0.320977)
(0.451325, 49.5122) +- (0.00334742, 0.301136)
(0.461991, 46.6103) +- (0.00310457, 0.2849769)
(0.472415, 44.1584) +- (0.0031121, 0.2724756)
(0.481311, 42.2831) +- (0.00309301, 0.2600767)
};
\addlegendentry{filtered-graphs algorithm, mean}
\addplot[dotted, mybars]
coordinates {
(0.134342, 104338.65) +- (0.00132285, 25084.626)
(0.195808, 63901.9) +- (0.00155259, 8700.546)
(0.239173, 47256.69) +- (0.00200563, 8541.224)
(0.26841, 37820.61) +- (0.00204706, 6990.608)
(0.292037, 24800.805) +- (0.00259634, 3102.3839)
(0.30921, 20422.937) +- (0.00310489, 2937.87)
(0.324333, 17485.114) +- (0.00298127, 3867.6563)
(0.354978, 9423.403) +- (0.00347603, 1175.2966)
(0.384942, 5368.328) +- (0.00361235, 688.6512)
(0.398999, 4423.165) +- (0.00345939, 671.9529)
(0.40933, 3531.406) +- (0.00336425, 484.2531)
(0.416517, 2936.454) +- (0.0034473, 391.313)
(0.436654, 2097.9928) +- (0.00333773, 247.45795)
(0.451325, 1596.9138) +- (0.00334742, 158.44872)
(0.461991, 1325.4151) +- (0.00310457, 135.18525)
(0.472415, 1138.7756) +- (0.0031121, 118.74796)
(0.481311, 1000.8354) +- (0.00309301, 101.54524)
};
\addlegendentry{brute-force algorithm, mean}
\addplot[thick, solid]
coordinates {
(0.134342, 6325.0) +- (0.00132285, 0)
(0.195808, 8330.0) +- (0.00155259, 0)
(0.239173, 8335.0) +- (0.00200563, 0)
(0.26841, 7395.0) +- (0.00204706, 0)
(0.292037, 7490.0) +- (0.00259634, 0)
(0.30921, 8575.0) +- (0.00310489, 0)
(0.324333, 7400.0) +- (0.00298127, 0)
(0.354978, 7885.0) +- (0.00347603, 0)
(0.384942, 7690.0) +- (0.00361235, 0)
(0.398999, 8165.0) +- (0.00345939, 0)
(0.40933, 8690.0) +- (0.00336425, 0)
(0.416517, 7195.0) +- (0.0034473, 0)
(0.436654, 6830.0) +- (0.00333773, 0)
(0.451325, 7040.0) +- (0.00334742, 0)
(0.461991, 6815.0) +- (0.00310457, 0)
(0.472415, 8315.0) +- (0.0031121, 0)
(0.481311, 9975.0) +- (0.00309301, 0)
};
\addlegendentry{proposed algorithm, max}
\addplot[thick, dashed]
coordinates {
(0.134342, 287.0) +- (0.00132285, 0)
(0.195808, 284.0) +- (0.00155259, 0)
(0.239173, 284.0) +- (0.00200563, 0)
(0.26841, 284.0) +- (0.00204706, 0)
(0.292037, 284.0) +- (0.00259634, 0)
(0.30921, 284.0) +- (0.00310489, 0)
(0.324333, 284.0) +- (0.00298127, 0)
(0.354978, 284.0) +- (0.00347603, 0)
(0.384942, 284.0) +- (0.00361235, 0)
(0.398999, 284.0) +- (0.00345939, 0)
(0.40933, 284.0) +- (0.00336425, 0)
(0.416517, 284.0) +- (0.0034473, 0)
(0.436654, 284.0) +- (0.00333773, 0)
(0.451325, 284.0) +- (0.00334742, 0)
(0.461991, 284.0) +- (0.00310457, 0)
(0.472415, 284.0) +- (0.0031121, 0)
(0.481311, 284.0) +- (0.00309301, 0)
};
\addlegendentry{filtered-graphs algorithm, max}
\addplot[thick, dotted]
coordinates {
(0.134342, 370067080.0) +- (0.00132285, 0)
(0.195808, 97594720.0) +- (0.00155259, 0)
(0.239173, 95788550.0) +- (0.00200563, 0)
(0.26841, 146809640.0) +- (0.00204706, 0)
(0.292037, 23440707.0) +- (0.00259634, 0)
(0.30921, 88895280.0) +- (0.00310489, 0)
(0.324333, 177317110.0) +- (0.00298127, 0)
(0.354978, 52675460.0) +- (0.00347603, 0)
(0.384942, 30470394.0) +- (0.00361235, 0)
(0.398999, 65633490.0) +- (0.00345939, 0)
(0.40933, 55849630.0) +- (0.00336425, 0)
(0.416517, 56039290.0) +- (0.0034473, 0)
(0.436654, 26184208.0) +- (0.00333773, 0)
(0.451325, 23011728.0) +- (0.00334742, 0)
(0.461991, 23011728.0) +- (0.00310457, 0)
(0.472415, 23011728.0) +- (0.0031121, 0)
(0.481311, 23011508.0) +- (0.00309301, 0)
};
\addlegendentry{brute-force algorithm, max}
\end{semilogyaxis}
\end{tikzpicture}}\hfill%
  \subfloat[memory used for $\gamma = 10$, and 640 units]{%
    \label{f:memory_10_640}%
    \begin{tikzpicture}
\begin{semilogyaxis}[xlabel = {network utilization}, ylabel = {memory words}, xmin = 0.1, xmax = 0.7, height = 3.8 cm, width = 6 cm, xlabel shift = -3 pt, legend columns = 3, legend to name = regular, legend style = {/tikz/every even column/.append style = {column sep = 0.25 cm}}, ymin = 1e1, ymax = 1e11, ylabel shift = -4 pt]
\addplot[solid]
coordinates {
(0.124004, 2038.122) +- (0.00104124, 41.40272)
(0.173418, 2268.092) +- (0.00153556, 48.4725)
(0.224284, 2245.015) +- (0.00173251, 55.2371)
(0.26456, 2151.452) +- (0.00187045, 51.9258)
(0.295828, 2015.377) +- (0.00261182, 41.27525)
(0.319363, 1889.37) +- (0.00275961, 36.01048)
(0.335841, 1787.22) +- (0.00306959, 33.31898)
(0.373713, 1554.345) +- (0.0033594, 27.0365)
(0.404553, 1349.6) +- (0.00343917, 21.74583)
(0.418391, 1262.35) +- (0.00357912, 19.32342)
(0.429609, 1194.525) +- (0.00347659, 17.52322)
(0.439332, 1135.755) +- (0.00348815, 16.73953)
(0.458518, 1025.405) +- (0.00340261, 15.37034)
(0.473956, 940.41) +- (0.00341648, 13.40917)
(0.486677, 874.56) +- (0.00318962, 11.61358)
(0.497379, 821.035) +- (0.00328742, 10.79715)
(0.506702, 776.783) +- (0.00319273, 10.11985)
};
\addlegendentry{proposed algorithm, mean}
\addplot[dashed]
coordinates {
(0.124004, 136.6492) +- (0.00104124, 0.590267)
(0.173418, 129.0564) +- (0.00153556, 0.588726)
(0.224284, 119.0942) +- (0.00173251, 0.612527)
(0.26456, 110.8761) +- (0.00187045, 0.565329)
(0.295828, 103.7303) +- (0.00261182, 0.474798)
(0.319363, 97.8028) +- (0.00275961, 0.393463)
(0.335841, 92.8408) +- (0.00306959, 0.399658)
(0.373713, 80.6784) +- (0.0033594, 0.37036)
(0.404553, 70.1957) +- (0.00343917, 0.378738)
(0.418391, 65.6921) +- (0.00357912, 0.355454)
(0.429609, 62.1511) +- (0.00347659, 0.364309)
(0.439332, 59.2609) +- (0.00348815, 0.349338)
(0.458518, 53.8235) +- (0.00340261, 0.325155)
(0.473956, 49.793) +- (0.00341648, 0.322835)
(0.486677, 46.7456) +- (0.00318962, 0.2951244)
(0.497379, 44.3356) +- (0.00328742, 0.2858265)
(0.506702, 42.3094) +- (0.00319273, 0.2856848)
};
\addlegendentry{filtered-graphs algorithm, mean}
\addplot[dotted, mybars]
coordinates {
(0.124004, 112996.65) +- (0.00104124, 14659.276)
(0.173418, 83953.55) +- (0.00153556, 9929.629)
(0.224284, 61354.74) +- (0.00173251, 7881.48)
(0.26456, 42759.56) +- (0.00187045, 5499.971)
(0.295828, 29514.66) +- (0.00261182, 3276.142)
(0.319363, 23644.108) +- (0.00275961, 3917.151)
(0.335841, 17499.445) +- (0.00306959, 1777.6639)
(0.373713, 9961.203) +- (0.0033594, 1190.4832)
(0.404553, 5733.577) +- (0.00343917, 679.3636)
(0.418391, 4635.482) +- (0.00357912, 605.5691)
(0.429609, 3749.465) +- (0.00347659, 494.4152)
(0.439332, 3156.315) +- (0.00348815, 372.8972)
(0.458518, 2440.202) +- (0.00340261, 335.25856)
(0.473956, 1910.2825) +- (0.00341648, 239.8592)
(0.486677, 1623.7398) +- (0.00318962, 226.54692)
(0.497379, 1406.4353) +- (0.00328742, 226.21117)
(0.506702, 1246.836) +- (0.00319273, 197.18683)
};
\addlegendentry{brute-force algorithm, mean}
\addplot[thick, solid]
coordinates {
(0.124004, 11735.0) +- (0.00104124, 0)
(0.173418, 14380.0) +- (0.00153556, 0)
(0.224284, 13625.0) +- (0.00173251, 0)
(0.26456, 14230.0) +- (0.00187045, 0)
(0.295828, 14785.0) +- (0.00261182, 0)
(0.319363, 14215.0) +- (0.00275961, 0)
(0.335841, 13880.0) +- (0.00306959, 0)
(0.373713, 12830.0) +- (0.0033594, 0)
(0.404553, 13665.0) +- (0.00343917, 0)
(0.418391, 13185.0) +- (0.00357912, 0)
(0.429609, 13220.0) +- (0.00347659, 0)
(0.439332, 13565.0) +- (0.00348815, 0)
(0.458518, 16205.0) +- (0.00340261, 0)
(0.473956, 18610.0) +- (0.00341648, 0)
(0.486677, 17715.0) +- (0.00318962, 0)
(0.497379, 15500.0) +- (0.00328742, 0)
(0.506702, 14495.0) +- (0.00319273, 0)
};
\addlegendentry{proposed algorithm, max}
\addplot[thick, dashed]
coordinates {
(0.124004, 290.0) +- (0.00104124, 0)
(0.173418, 287.0) +- (0.00153556, 0)
(0.224284, 287.0) +- (0.00173251, 0)
(0.26456, 287.0) +- (0.00187045, 0)
(0.295828, 287.0) +- (0.00261182, 0)
(0.319363, 287.0) +- (0.00275961, 0)
(0.335841, 287.0) +- (0.00306959, 0)
(0.373713, 284.0) +- (0.0033594, 0)
(0.404553, 284.0) +- (0.00343917, 0)
(0.418391, 284.0) +- (0.00357912, 0)
(0.429609, 284.0) +- (0.00347659, 0)
(0.439332, 284.0) +- (0.00348815, 0)
(0.458518, 284.0) +- (0.00340261, 0)
(0.473956, 284.0) +- (0.00341648, 0)
(0.486677, 284.0) +- (0.00318962, 0)
(0.497379, 284.0) +- (0.00328742, 0)
(0.506702, 284.0) +- (0.00319273, 0)
};
\addlegendentry{filtered-graphs algorithm, max}
\addplot[thick, dotted]
coordinates {
(0.124004, 89577920.0) +- (0.00104124, 0)
(0.173418, 53784960.0) +- (0.00153556, 0)
(0.224284, 63344010.0) +- (0.00173251, 0)
(0.26456, 82149860.0) +- (0.00187045, 0)
(0.295828, 36722260.0) +- (0.00261182, 0)
(0.319363, 302292990.0) +- (0.00275961, 0)
(0.335841, 36392190.0) +- (0.00306959, 0)
(0.373713, 69923410.0) +- (0.0033594, 0)
(0.404553, 36383120.0) +- (0.00343917, 0)
(0.418391, 56120000.0) +- (0.00357912, 0)
(0.429609, 36380980.0) +- (0.00347659, 0)
(0.439332, 49669890.0) +- (0.00348815, 0)
(0.458518, 161183850.0) +- (0.00340261, 0)
(0.473956, 119978650.0) +- (0.00341648, 0)
(0.486677, 119978650.0) +- (0.00318962, 0)
(0.497379, 119978650.0) +- (0.00328742, 0)
(0.506702, 119978650.0) +- (0.00319273, 0)
};
\addlegendentry{brute-force algorithm, max}
\end{semilogyaxis}
\end{tikzpicture}}\\[10pt]
  \centering\ref{regular}%
  \caption{Simulation results: the sample means and maxima of the time taken and memory used by the evaluated algorithms.}
  \label{f:plots}
\end{figure*}

Figure \ref{f:plots} shows the sample means and the sample maxima of
the time taken and memory used by a shortest path search, regardless
of whether the search was successful or not.  The results are shown on
a logarithmic scale as a function of network utilization.  The curves
are plotted solid for the proposed algorithm, dashed for the
filtered-graphs algorithm, and dotted for the brute-force algorithm.
The sample means are plotted thin, and the sample maxima thick.  Each
curve is drawn using 17 data points for different values of $\mu$.
The error bars of the 95\% confidence interval, appear only for the
sample means of the brute-force algorithm, since their relative
standard error was high, frequently around 20\%, while for the other
sample means the error bars were too small to plot.

% The subfigures.

Figure \ref{f:plots} has 12 subfigures in four rows and three columns.
The first and the second rows show the time results for $\gamma = 1$,
and $\gamma = 10$, respectively, and the third and fourth rows show
the memory results for $\gamma = 1$, and $\gamma = 10$, respectively.
The first column shows the results for $|\Omega| = 160$, the second
for $|\Omega| = 320$, and the third for $|\Omega| = 640$.  We use the
same scales in the various plots to allow for easy comparison.

% The sample mean time results for $\gamma = 1$.

The sample mean time results for $\gamma = 1$ show that the proposed
algorithm was usually about 10 times faster and at most 20 times
faster than two other algorithms, except at very heavy utilization,
where most searches ended up with no solution, and where the
brute-force algorithm was able to determine this more quickly.  As the
utilization increased, the mean time of every algorithm decreased, as
the solution was less likely to exist.  As the number of units
increased from 160 to 320, and from 320 to 640 the mean time results
increased about twice for the proposed algorithm and the
filtered-graphs algorithm, and stayed about the same for the
brute-force algorithm.

% The sample mean time results for $\gamma = 10$.

The sample mean time results for $\gamma = 10$ show that the proposed
algorithm \emph{was hundreds of times} faster than two other
algorithms, and for the case of 640 units and light utilization, the
proposed algorithm was about 500 times faster than the filtered-graphs
algorithm.

% The pessimistic performance.

As to the pessimistic time performance, i.e., the sample maximum time
taken, our algorithm and the filtered-graphs algorithm usually took a
few seconds, while the brute-force algorithm usually took hundreds of
seconds, which makes a gap of \emph{two orders of magnitude}.

% The memory results.

The memory results are clear-cut: the filtered-graphs algorithm
performed the best (since it used the Dijkstra algorithm), our
algorithm performed very well, and the brute-force algorithm performed
the worst.  While the sample mean memory results of the brute-force
algorithm are acceptable (even better than those of the proposed
algorithm for heavy utilization), the sample maximum memory results
reveal the unacceptable memory performance of the brute-force
algorithm: the brute-force algorithm required about \emph{six orders
of magnitude} more memory than the proposed and the filtered-graphs
algorithms.

% Changes in the memory used for different \gamma and units.

As the number of units increased from 160 to 320, and from 320 to 640
the mean memory results increased almost twice for our algorithm, for
the filtered-graphs algorithm the results expectedly stayed the same,
and for the brute-force algorithm increased about 10\%.  The memory
used by our algorithm and the brute-force algorithm were smaller for
$\gamma = 10$ in comparison with $\gamma = 1$, because the spectrum
was fragmented less.

% The stacked memory plots.

\begin{figure*}
  \subfloat[Proposed algorithm.]{%
    \label{f:stack_dijkstra}%
    \begin{tikzpicture}
\begin{axis}[xlabel = {network utilization}, ylabel = {memory words}, stack plots = y, area style, enlarge x limits = false, height = 3.8 cm, width = 6 cm, ymin = 0, xlabel shift = -3 pt, ylabel shift = -3 pt, legend columns = -1, legend to name = stack, legend style = {/tikz/every even column/.append style = {column sep = 0.25 cm}}, scaled y ticks = base 10:-4]
\addplot[pattern = north east lines]
coordinates { (0.127142, 5613.0) +- (0.000588861, 0)
(0.19061, 6903.0) +- (0.000927054, 0)
(0.253873, 7255.0) +- (0.00161753, 0)
(0.320806, 6799.0) +- (0.00220855, 0)
(0.370566, 6088.0) +- (0.00265177, 0)
(0.401078, 5984.0) +- (0.00336707, 0)
(0.421552, 5969.0) +- (0.00382383, 0)
(0.469919, 6425.0) +- (0.00453692, 0)
(0.50895, 6164.0) +- (0.00480454, 0)
(0.526206, 6191.0) +- (0.00483655, 0)
(0.53985, 6264.0) +- (0.00486905, 0)
(0.551682, 5598.0) +- (0.00476024, 0)
(0.574532, 6145.0) +- (0.00456749, 0)
(0.59286, 5441.0) +- (0.00435369, 0)
(0.608082, 5757.0) +- (0.00420392, 0)
(0.621302, 5802.0) +- (0.00412613, 0)
(0.632771, 5967.0) +- (0.00406718, 0)
}\closedcycle;
\addlegendentry{costs}
\addplot[pattern = crosshatch]
coordinates { (0.127142, 11226.0) +- (0.000588861, 0)
(0.19061, 13806.0) +- (0.000927054, 0)
(0.253873, 14510.0) +- (0.00161753, 0)
(0.320806, 13598.0) +- (0.00220855, 0)
(0.370566, 12176.0) +- (0.00265177, 0)
(0.401078, 11968.0) +- (0.00336707, 0)
(0.421552, 11938.0) +- (0.00382383, 0)
(0.469919, 12850.0) +- (0.00453692, 0)
(0.50895, 12328.0) +- (0.00480454, 0)
(0.526206, 12382.0) +- (0.00483655, 0)
(0.53985, 12528.0) +- (0.00486905, 0)
(0.551682, 11196.0) +- (0.00476024, 0)
(0.574532, 12290.0) +- (0.00456749, 0)
(0.59286, 10882.0) +- (0.00435369, 0)
(0.608082, 11514.0) +- (0.00420392, 0)
(0.621302, 11604.0) +- (0.00412613, 0)
(0.632771, 11934.0) +- (0.00406718, 0)
}\closedcycle;
\addlegendentry{edges}
\addplot[pattern = north west lines]
coordinates { (0.127142, 11226.0) +- (0.000588861, 0)
(0.19061, 13806.0) +- (0.000927054, 0)
(0.253873, 14510.0) +- (0.00161753, 0)
(0.320806, 13598.0) +- (0.00220855, 0)
(0.370566, 12176.0) +- (0.00265177, 0)
(0.401078, 11968.0) +- (0.00336707, 0)
(0.421552, 11938.0) +- (0.00382383, 0)
(0.469919, 12850.0) +- (0.00453692, 0)
(0.50895, 12328.0) +- (0.00480454, 0)
(0.526206, 12382.0) +- (0.00483655, 0)
(0.53985, 12528.0) +- (0.00486905, 0)
(0.551682, 11196.0) +- (0.00476024, 0)
(0.574532, 12290.0) +- (0.00456749, 0)
(0.59286, 10882.0) +- (0.00435369, 0)
(0.608082, 11514.0) +- (0.00420392, 0)
(0.621302, 11604.0) +- (0.00412613, 0)
(0.632771, 11934.0) +- (0.00406718, 0)
}\closedcycle;
\addlegendentry{units}
\end{axis}
\end{tikzpicture}}\hfill%
  \subfloat[Filtered-graphs algorithm.]{%
    \label{f:stack_parallel}%
    \begin{tikzpicture}
\begin{axis}[xlabel = {network utilization}, ylabel = {memory words}, stack plots = y, area style, enlarge x limits = false, height = 3.8 cm, width = 6 cm, ymin = 0, xlabel shift = -3 pt, ylabel shift = -3 pt, legend columns = -1, legend to name = stack, legend style = {/tikz/every even column/.append style = {column sep = 0.25 cm}}, scaled y ticks = base 10:-2]
\addplot[pattern = north east lines]
coordinates { (0.127142, 98.0) +- (0.000588861, 0)
(0.19061, 98.0) +- (0.000927054, 0)
(0.253873, 98.0) +- (0.00161753, 0)
(0.320806, 98.0) +- (0.00220855, 0)
(0.370566, 98.0) +- (0.00265177, 0)
(0.401078, 98.0) +- (0.00336707, 0)
(0.421552, 98.0) +- (0.00382383, 0)
(0.469919, 96.0) +- (0.00453692, 0)
(0.50895, 96.0) +- (0.00480454, 0)
(0.526206, 96.0) +- (0.00483655, 0)
(0.53985, 96.0) +- (0.00486905, 0)
(0.551682, 96.0) +- (0.00476024, 0)
(0.574532, 96.0) +- (0.00456749, 0)
(0.59286, 96.0) +- (0.00435369, 0)
(0.608082, 96.0) +- (0.00420392, 0)
(0.621302, 96.0) +- (0.00412613, 0)
(0.632771, 96.0) +- (0.00406718, 0)
}\closedcycle;
\addlegendentry{costs}
\addplot[pattern = crosshatch]
coordinates { (0.127142, 196.0) +- (0.000588861, 0)
(0.19061, 196.0) +- (0.000927054, 0)
(0.253873, 196.0) +- (0.00161753, 0)
(0.320806, 196.0) +- (0.00220855, 0)
(0.370566, 196.0) +- (0.00265177, 0)
(0.401078, 196.0) +- (0.00336707, 0)
(0.421552, 196.0) +- (0.00382383, 0)
(0.469919, 192.0) +- (0.00453692, 0)
(0.50895, 192.0) +- (0.00480454, 0)
(0.526206, 192.0) +- (0.00483655, 0)
(0.53985, 192.0) +- (0.00486905, 0)
(0.551682, 192.0) +- (0.00476024, 0)
(0.574532, 192.0) +- (0.00456749, 0)
(0.59286, 192.0) +- (0.00435369, 0)
(0.608082, 192.0) +- (0.00420392, 0)
(0.621302, 192.0) +- (0.00412613, 0)
(0.632771, 192.0) +- (0.00406718, 0)
}\closedcycle;
\addlegendentry{edges}
\addplot[pattern = north west lines]
coordinates { (0.127142, 2.0) +- (0.000588861, 0)
(0.19061, 2.0) +- (0.000927054, 0)
(0.253873, 2.0) +- (0.00161753, 0)
(0.320806, 2.0) +- (0.00220855, 0)
(0.370566, 2.0) +- (0.00265177, 0)
(0.401078, 2.0) +- (0.00336707, 0)
(0.421552, 2.0) +- (0.00382383, 0)
(0.469919, 2.0) +- (0.00453692, 0)
(0.50895, 2.0) +- (0.00480454, 0)
(0.526206, 2.0) +- (0.00483655, 0)
(0.53985, 2.0) +- (0.00486905, 0)
(0.551682, 2.0) +- (0.00476024, 0)
(0.574532, 2.0) +- (0.00456749, 0)
(0.59286, 2.0) +- (0.00435369, 0)
(0.608082, 2.0) +- (0.00420392, 0)
(0.621302, 2.0) +- (0.00412613, 0)
(0.632771, 2.0) +- (0.00406718, 0)
}\closedcycle;
\addlegendentry{units}
\end{axis}
\end{tikzpicture}}\hfill%
  \subfloat[Brute-force algorithm.]{%
    \label{f:stack_brtforce}%
    \begin{tikzpicture}
\begin{axis}[xlabel = {network utilization}, ylabel = {memory words}, stack plots = y, area style, enlarge x limits = false, height = 3.8 cm, width = 6 cm, ymin = 0, xlabel shift = -3 pt, ylabel shift = -3 pt, legend columns = -1, legend to name = stack, legend style = {/tikz/every even column/.append style = {column sep = 0.25 cm}}, ]
\addplot[pattern = north east lines]
coordinates { (0.127142, 1372880.0) +- (0.000588861, 0)
(0.19061, 2895130.0) +- (0.000927054, 0)
(0.253873, 49206300.0) +- (0.00161753, 0)
(0.320806, 4090090.0) +- (0.00220855, 0)
(0.370566, 6485640.0) +- (0.00265177, 0)
(0.401078, 7088280.0) +- (0.00336707, 0)
(0.421552, 45738000.0) +- (0.00382383, 0)
(0.469919, 26399200.0) +- (0.00453692, 0)
(0.50895, 4875460.0) +- (0.00480454, 0)
(0.526206, 5019120.0) +- (0.00483655, 0)
(0.53985, 19101300.0) +- (0.00486905, 0)
(0.551682, 3579090.0) +- (0.00476024, 0)
(0.574532, 9312280.0) +- (0.00456749, 0)
(0.59286, 11662500.0) +- (0.00435369, 0)
(0.608082, 3616930.0) +- (0.00420392, 0)
(0.621302, 9335340.0) +- (0.00412613, 0)
(0.632771, 9796830.0) +- (0.00406718, 0)
}\closedcycle;
\addlegendentry{costs}
\addplot[pattern = crosshatch]
coordinates { (0.127142, 14280400.0) +- (0.000588861, 0)
(0.19061, 16670100.0) +- (0.000927054, 0)
(0.253873, 137588000.0) +- (0.00161753, 0)
(0.320806, 12068700.0) +- (0.00220855, 0)
(0.370566, 14638600.0) +- (0.00265177, 0)
(0.401078, 25172900.0) +- (0.00336707, 0)
(0.421552, 119070000.0) +- (0.00382383, 0)
(0.469919, 54496800.0) +- (0.00453692, 0)
(0.50895, 20097200.0) +- (0.00480454, 0)
(0.526206, 18517100.0) +- (0.00483655, 0)
(0.53985, 44591900.0) +- (0.00486905, 0)
(0.551682, 16348500.0) +- (0.00476024, 0)
(0.574532, 28468500.0) +- (0.00456749, 0)
(0.59286, 40180300.0) +- (0.00435369, 0)
(0.608082, 16660600.0) +- (0.00420392, 0)
(0.621302, 52559000.0) +- (0.00412613, 0)
(0.632771, 45923800.0) +- (0.00406718, 0)
}\closedcycle;
\addlegendentry{edges}
\addplot[pattern = north west lines]
coordinates { (0.127142, 47077400.0) +- (0.000588861, 0)
(0.19061, 108748000.0) +- (0.000927054, 0)
(0.253873, 2143620000.0) +- (0.00161753, 0)
(0.320806, 167214000.0) +- (0.00220855, 0)
(0.370566, 249679000.0) +- (0.00265177, 0)
(0.401078, 287249000.0) +- (0.00336707, 0)
(0.421552, 1978720000.0) +- (0.00382383, 0)
(0.469919, 1145870000.0) +- (0.00453692, 0)
(0.50895, 195095000.0) +- (0.00480454, 0)
(0.526206, 171567000.0) +- (0.00483655, 0)
(0.53985, 810208000.0) +- (0.00486905, 0)
(0.551682, 162265000.0) +- (0.00476024, 0)
(0.574532, 360654000.0) +- (0.00456749, 0)
(0.59286, 462266000.0) +- (0.00435369, 0)
(0.608082, 137606000.0) +- (0.00420392, 0)
(0.621302, 319682000.0) +- (0.00412613, 0)
(0.632771, 334638000.0) +- (0.00406718, 0)
}\closedcycle;
\addlegendentry{units}
\end{axis}
\end{tikzpicture}}\\[10pt]
  \centering\ref{stack}%
  \caption{Simulation results: the maximum number of required words by
    the evaluated algorithms.}
  \label{f:stack}
\end{figure*}
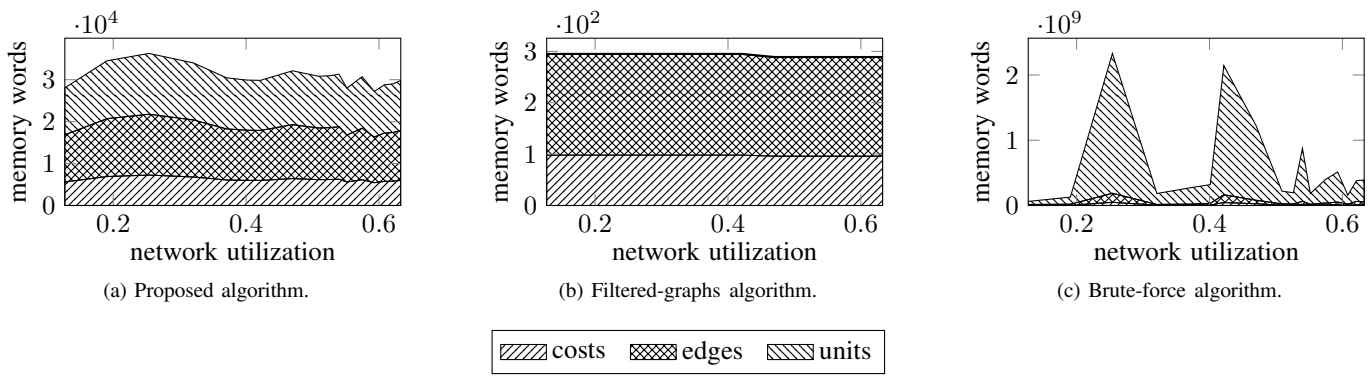

Figure \ref{f:stack} shows the stacked plots of the maximum number of
required memory words by our algorithm (Fig.~\ref{f:stack_dijkstra}),
the filtered-graphs algorithm (Fig.~\ref{f:stack_parallel}), and the
brute-force algorithm (Fig.~\ref{f:stack_brtforce}).  In each of the
plots there are 17 $\times$ 3 data points, because for the 17 values
of the offered load $\mu$ we report the maximum number of required
costs, edges, and units.  The results are at the order of $10^4$ for
our algorithm, $10^2$ for the filtered-graphs algorithm, and $10^9$
for the brute-force algorithm.  Clearly, the filtered-graphs algorithm
performs best (requiring at most 1.2 kB), followed by our algorithm
(requiring at most 160 kB), and then followed by the brute-force
algorithm (requiring at most 10 GB).  For our algorithm, the memory
was used in 20\% for costs, in 40\% for edges, and 40\% for units,
since a label has one cost (one word), one edge (two words), and one
CU (two words).  For the filtered-graphs algorithm, the memory was
used mainly to store edges, and costs, since a label has one cost and
one edge.  The brute-force algorithm used most of its memory to store
the units of the paths in the priority queue.

Table \ref{t:summary} summarizes the algorithm comparison.

\begin{table}
  \caption{Summary of algorithm comparison.}
  \label{t:summary}
  \centering
  \begin{tabular}{|p{0.21\columnwidth}|p{0.325\columnwidth}|p{0.325\columnwidth}|}
    \hline
    \textbf{Algorithm} & \textbf{Pros} & \textbf{Cons}\\
    \hline
    generic Dijkstra & fastest & complex\\
    \hline
    filtered-graphs & ultra-low memory usage & slow\\
    \hline
    brute-force & good for small networks & fails for large networks\\
    \hline
  \end{tabular}
\end{table}

%%%%%%%%%%%%%%%%%%%%%%%%%%%%%%%%%%%%%%%%%%%%%%%%%%%%%%%%%%%%%%%%%%%%%%%%%%%

\section{Conclusion}
\label{conclusion}

% The contribution in a nutshel.

We proposed a novel generalization of the Dijkstra shortest path
algorithm for finding a shortest path in the wavelength-division
multiplexed networks and the elastic optical networks.

% Simulation results.

Our extensive simulation studies show that the proposed algorithm has
a small memory footprint and is considerably (even hundreds of times)
faster than other routing algorithms that are frequently utilized.

% We are confident, though we have no proof.

We provide no proof of correctness or complexity analysis.  However,
with robust simulations, we corroborated the correctness and
efficiency of the proposed algorithm.

% Not only the optical network.

We presented the algorithm in the setting of optical networks, but we
believe that the novel ideas we introduced can be applied in the
routing in the multilayer and wireless networks to make their control
quicker.  For instance, the algorithm could be used to solve
efficiently the contiguous frequency and time resource allocation in
the wireless orthogonal frequency-multiplexed wireless networks.

% Future work.

There are a number of directions for future work.  First, the
algorithm could be adapted for parallel execution, making it even
faster.  Second, the algorithm could be turned into a distributed
algorithm, very much like a distance-vector algorithm.  Next, the
algorithm could be extended further to be applicable to multilayer
networks or space-division multiplexed networks.  Finally, the
algorithm could be extended to take into account signal regeneration,
spectrum conversion, or inverse multiplexing.

\section*{Acknowledgments}

I, Irek Szcześniak, dedicate this work to my mom and dad, Halina and
Jerzy Szcześniak, for their never-ending love and support.

We thank the anonymous reviewers for their constructive criticism,
which helped us improve our work.

\bibliography{article}

\end{document}